\documentclass[largeformat,final]{interact}
\usepackage[utf8]{inputenc}
\usepackage[T1]{fontenc}
\usepackage{lscape}
\usepackage{subfiles} 
\usepackage{comment}
\usepackage{tabularx}
\usepackage{graphicx}
\usepackage{textcomp}
\usepackage{lmodern}
\usepackage{url}
\usepackage{multirow}
\usepackage{subfigure}
\usepackage[hidelinks]{hyperref}
\usepackage{orcidlink}

\usepackage[numbers,sort&compress]{natbib} 
\bibpunct[, ]{[}{]}{,}{n}{,}{,} 

\theoremstyle{plain} 

\theoremstyle{definition}

\theoremstyle{remark}

\begin{document}

\title{Establishing assembly-oriented modular product architectures through Design for Assembly enhanced Modular Function Deployment}

\author{
\name{Fabio Marco Monetti\orcidlink{0000-0003-2993-511X}\textsuperscript{a}$^\ast$\thanks{$^\ast$CONTACT Fabio Marco Monetti. Email: monetti@kth.se}, Adam Lundström\orcidlink{0009-0008-8432-9167}\textsuperscript{a,b}, Colin de Kwant\orcidlink{0000-0001-8692-5589}\textsuperscript{b}, Magnus Gyllenskepp\textsuperscript{b} and Antonio Maffei\orcidlink{0000-0002-0723-1712}\textsuperscript{a}}
\affil{\textsuperscript{a}KTH Royal Institute of Technology, Brinellvägen 68, 114 28, Stockholm, SE; \textsuperscript{b}Sweden Modular Management AB, Kungsgatan 56, 111 22, Stockholm, SE}
}

\maketitle

\begin{abstract}
Modular product design has become a strategic enabler for companies seeking to balance product variety, operational efficiency, and market responsiveness, making the alignment between modular architecture and manufacturing considerations increasingly critical. Modular Function Deployment (MFD) is a widely adopted method for defining modular product architectures, yet it lacks systematic support for assembly considerations during early concept and system-level development. This limitation increases the risk of delayed production ramp-up and lifecycle inefficiencies.
This paper proposes a set of enhancements to MFD that integrate Design for Assembly (DFA) logic into architectural synthesis. The extended method introduces structured heuristics, assembly-oriented module drivers, a coded interface taxonomy, and quantitative metrics for assessing assembly feasibility and automation readiness. These additions preserve compatibility with standard MFD workflows while enriching decision-making with traceable, production-informed reasoning.
An illustrative case study involving a handheld leaf blower demonstrates the method’s usability and effectiveness. The redesigned architecture shows reduced assembly effort, simplified interfaces, and increased automation potential. By supporting early-stage evaluation of architectural alternatives through an assembly lens, the method enables faster transition to efficient volume production and provides a foundation for continuous improvement throughout the product lifecycle.
\end{abstract}


\section{Introduction} \label{sec:intro}
Product design and development are shaped by a company’s strategic positioning, guiding resource allocation, and long-term competitiveness. Aligning design efforts with business strategies helps manage complexity while meeting customer expectations~\cite{treacy_customer_1993, miller_taxonomy_1994}. Such strategies are a convergence of market requirements and internal operational constraints that shape how products are industrialised~\cite{frohlich_taxonomy_2001}.

Mass customisation has made modular product design a strategic stabiliser of complexity, production efficiency, and variety~\cite{pine_mass_1993, ulrich_role_1995}, increasing the impact of architectural decisions on downstream activities. Beyond engineering value, modularisation is a broader business tool:~it supports cost reduction, operational flexibility, and alignment with sustainability goals~\cite{fixson_modularity_2007,larsson_standardization_2018}. For instance, Scania’s modular platform improved manufacturing efficiency while enabling product variety~\cite{nilsson_strategic_1995}.

As modular strategies mature, aligning architectural decisions with manufacturing and assembly constraints becomes critical to realise their full operational potential. However, early design decisions are often made without consideration of how they affect assembly feasibility, ergonomics, automation, or supply-chain~\cite{gershenson_product_2003,howard_modularization_2007,abdoli_modelling_2023}. This lack can delay ramp-up, increase rework, and lead to unrecoverable losses in profit over the product life~\cite{ulrich_product_2020}.

Modular Function Deployment (MFD) is a modularisation method to manage product variety and complexity. MFD supports early structuring by linking customer needs to technical solutions (TS) and grouping functions into coherent modules~\cite{erixon_modular_1998}. However, while MFD excels at managing functional and market-driven variation~\cite{ericsson_controlling_1999}, it often postpones the consideration of production-related factors. Thus, architectural decisions can be made without a clear understanding of their implications~\cite{favi_method_2012}.

Design for Assembly (DFA) provides structured guidance to simplify products and cut assembly cost by analysing part count, handling, insertion, and automation readiness.~\cite{boothroyd_product_2010}. Yet, DFA is typically applied only after core architectural decisions are locked in~\cite{favi_method_2012}, and its embodiment-level assessment often limits its influence on strategic modularisation choices, where bigger leverage on product lifecycle cost and manufacturability exists~\cite{eskilander_design_2001}.

Meanwhile, sustainability and Circular Economy (CE) agendas have renewed focus on lifecycle strategies like reuse, remanufacturing, and end-of-life disassembly~\cite{den_hollander_product_2017}. Modularisation plays a critical role here, and many principles that help assembly are equally beneficial for disassembly~\cite{mesa_linking_2022}. Although centred on assembly, the proposed method can extend to CE objectives, underscoring the need to embed lifecycle considerations in early architectural decisions.

This paper addresses the need for integrated architectural decisions linking customer-driven modularisation with production feasibility. It presents an expanded MFD method that embeds DFA logic into early concept evaluation and architecture definition, demonstrated on a handheld leaf blower by comparing a legacy configuration with a redesign optimised for assembly and automation.

Although developed for analysis, DFA is often used synthetically to refine designs by reducing parts or simplifying operations. This paper restores DFA’s analytical role by positioning it earlier in modular design, guiding decisions from the outset. We frame DFA as a lens on architectural feasibility, clarifying how design choices shape assembly performance.

The aim is to help teams evaluate and select concepts, module groupings, and interface strategies on both functional and feasibility grounds before geometry is fixed. Integrating assembly-aware criteria and structured complexity assessment enables more informed, traceable, and production-aligned architectural decisions.
\section{Background} \label{sec:background}
The product development process is shaped by early-stage design decisions, which determine product structure, cost, and manufacturing feasibility~\cite{hubka_theory_1988, pahl_engineering_2007, ulrich_product_2020}. In modular architectures, decisions such as module boundaries, functional allocation, and interface characteristics carry long-term implications for manufacturability~\cite{holtta-otto_degree_2007}. Since a substantial portion of lifecycle costs is committed here, later changes are typically disruptive and costly~\cite{blanchard_design_1978}. Consequently, identifying and refining modular concepts early is both a design challenge and a strategic priority.

Modularisation often prioritises market concerns:~customer segmentation, value delivery, and configurability~\cite{miller_taxonomy_1994}. While vital for variety and responsiveness, overemphasis can compromise production efficiency~\cite{elmaraghy_changeable_2009}. Assembly-oriented thinking remains under-represented in architecture, despite its importance for planning, standardisation, and system-level optimisation~\cite{boothroyd_product_2010}.

Modular product design is both technical and strategic, shaping how companies compete. Value disciplines~\cite{treacy_customer_1993} suggest firms orient strategy toward Operational Excellence, Customer Intimacy, or Product Leadership. Operational Excellence stresses efficiency, variant management, and supply chain rationalisation. Yet assembly feasibility and automation potential remain  soemhow neglected~\cite{monetti_barriers_2024}, exposing a gap between strategic intent and architectural decisions.

\subsection{Existing methods and attempts to integrate DFA into modular design}
Early DFA integration refers to inclusion of related considerations during the initial stages of design. Similarly, an assembly-oriented modular architecture describes a product structure designed with explicit regard for assembly constraints, manufacturability, and automation feasibility~\cite{monetti_towards_2023}. 

Over the years, a wide range of modularisation methods has been proposed, with functional decomposition, clustering of design interdependencies, and multi-objective optimisation. These offer limited support for the integration of DFA, particularly when architecture and interface definitions are still easily modified~\cite{favi_method_2012}.

Function-based methods define module boundaries via functional heuristics such as energy paths, conversion–transmission logic, and clustering~\cite{stone_development_2000}. They suit early stages when geometric or economic data are lacking~\cite{stone_product_2004}, but are hard to apply consistently across complex families with multiple alternatives. They rarely assess assembly effort or support formalised interface evaluation~\cite{favi_method_2012} and may overlook how cumulative part-level simplicity can increase overall complexity~\cite{rodriguez-toro_complexity_2003}.

Matrix-based methods, especially the Design Structure Matrix (DSM), map interdependencies among components, functions, or organisations~\cite{pimmler_integration_1994, browning_applying_2001}. DSM clustering helps reveal modules and simplify complexity. Yet standard DSMs ignore manufacturing factors such as handling, joining, or sequencing. The Module Interface Graph (MIG) improves representation by classifying interfaces~\cite{blees_development_2008}, but still offers limited DFA integration, overlooking constraints like accessibility or fixtures. Even enriched DSMs struggle with high-complexity systems due to indirect dependencies and ambiguous clustering~\cite{holmqvist_analysis_2003}.

Optimisation-based methods, e.g., TOPSIS~\cite{favi_multi-objective_2016} and Particle Swarm Optimisation (PSO)~\cite{chiu_redesign_2016}, balance multiple criteria effectively when quantitative data are available. However, they mainly assess pre-defined architectures rather than generate concepts. They also lack mechanisms for assembly-oriented design, such as defining interfaces, joining directions, or mating sequences. DFA constraints appear only after modules and materials are fixed, limiting upstream relevance.

The diversity of approaches and absence of a common vocabulary hinder the systematic integration of assembly as a driver for modularisation~\cite{bonvoisin_systematic_2016}. Key factors are seldom addressed generatively:~even recent modular evaluation methods still relegate DFA to a downstream scoring step~\cite{formentini_design_2022, dong_dfma-oriented_2023}, limiting effectiveness in production-sensitive contexts.

\begin{figure}
    \centering
    \includegraphics[width=\linewidth]{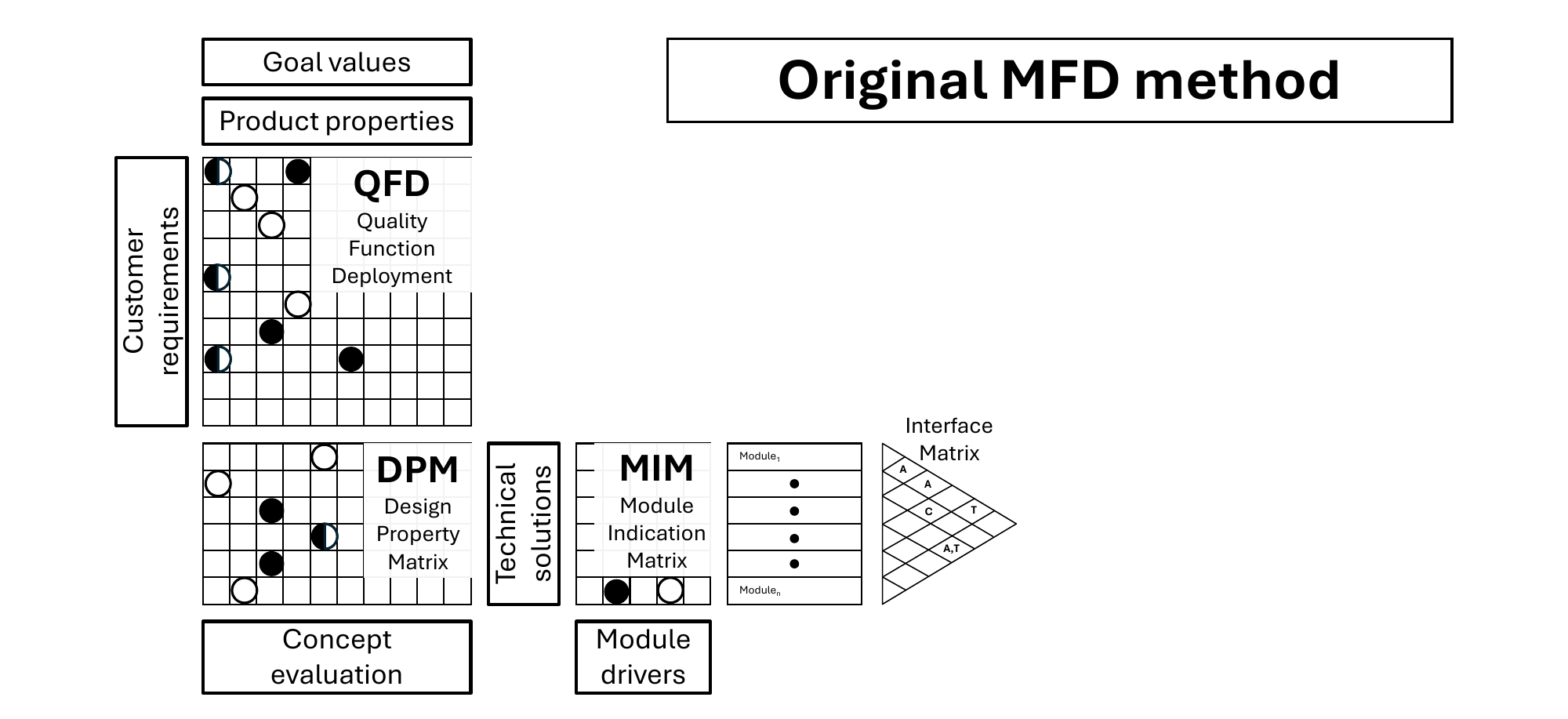}
    \caption{\textbf{Original MFD method (adapted from~\cite{lange_modular_2014}).} It links customer requirements to TS through a structured flow of matrices. Evaluation relies on functional and market-related criteria. \label{fig:mfd_original}}
\end{figure}

Even MFD, which aligns customer value, functions, and modular architecture through structured matrices~\cite{ericsson_controlling_1999}, does not inherently consider assembly (see Figure~\ref{fig:mfd_original}). Tools like the Design Property Matrix (DPM) and Module Indication Matrix (MIM) link properties, functions, and module candidates using Module Drivers (MD) like styling, carry-over, or serviceability, but overlook orientation, joining sequence, accessibility, and tool use.

DFA typically enters late, in the `Improve each module' step, applied synthetically after interfaces are fixed~\cite{erixon_modular_1998}. This constrains broader architectural changes. While MFD’s strength lies in bridging technical and commercial goals and supporting cross-functional work~\cite{holmqvist_analysis_2003, monetti_barriers_2024}, it does not systematically embed assembly logic. Still, its use has cut part count, lead time, and complexity in firms such as Husqvarna and MTS Systems.

Manufacturing concerns are often deferred to late stages, where DFX resolves local issues post-hoc~\cite{monetti_barriers_2024}. This delays system-level planning and wastes potential gains from automation compatibility, interface standardisation, and parallel assembly strategies.

Several methods attempt earlier integration. Assembly-Oriented Design (AOD) embeds constraints via model-based strategies~\cite{demoly_assembly_2011}, but often require complex tools. Conceptual Design for Assembly (cDFA) supports part count reduction using functional heuristics~\cite{stone_product_2004}; useful for ideation, but largely qualitative and not suited for architectural synthesis.

An extended cDFA enables structured, quantitative assessment of assembly at the conceptual stage~\cite{favi_method_2012, formentini_cdfa_2022}. Using hierarchical attributes, it highlights inefficiencies and supports more assembly-friendly designs, as shown in applications like aircraft nose–fuselage systems~\cite{formentini_method_2022}. Multi-criteria methods further help balance assembly with other design factors~\cite{favi_multi-objective_2016}.

Yet cDFA remains mainly diagnostic. It does not explicitly guide modular architectures or interface classification—critical for translating DFA insights into architectural choices. Its integration with modularisation is thus limited, especially where early trade-offs between function, assembly, and modularity must be visualised. This leaves room for complementary methods embedding assembly logic into architectural generation, interface planning, and early module definition.

Despite these efforts, a gap remains in analytically embedding DFA into architectural decision-making. Most approaches emphasise evaluation rather than guiding modular structures from an assembly standpoint. Table~\ref{tab:methods_dfa_comparison} summarises existing methods’ strengths and limitations from a production perspective.

\begin{table}
\tbl{Comparison of modularisation methods with respect to assembly integration.}
{\begin{tabular}{>{\raggedright\arraybackslash}p{0.20\linewidth}
>{\raggedright\arraybackslash}p{0.15\linewidth}
>{\raggedright\arraybackslash}p{0.25\linewidth}
>{\raggedright\arraybackslash}p{0.30\linewidth}}
\toprule
\textbf{Focus} & \textbf{Methods} & \textbf{Strengths} & \textbf{Assembly-related limitations} \\
\midrule

\textit{Modular architecture creation} & MFD~\cite{erixon_modular_1998, ericsson_controlling_1999} & Structured early-stage concept mapping and functional decomposition & No early assembly logic; lacks orientation, joining planning support \\
\addlinespace

\textit{Assembly heuristics-based design} & DFA, cDFA~\cite{boothroyd_product_2010,stone_product_2004,eskilander_design_2001,favi_method_2012} & Clear rules for simplification; concept-level DFA metrics & Typically post hoc; limited architectural or interface guidance \\
\addlinespace

\textit{Model-based assembly-oriented design} & AOD~\cite{demoly_assembly_2011} & Detailed CAD-based constraints and join simulation & High modelling overhead; limited scalability to conceptual phases \\
\addlinespace

\textit{System visualisation and dependency mapping} & DSM, MIG~\cite{pimmler_integration_1994, browning_applying_2001, blees_development_2008} & Visual interface clustering; functional dependencies & No generative logic; lacks DFA criteria; minimal guidance for interface simplification \\
\addlinespace

\textit{Optimisation-based modularisation} & TOPSIS, PSO~\cite{favi_multi-objective_2016, chiu_redesign_2016} & Multi-criteria balancing (cost, weight, DFA scores) & Requires complete input data; less suitable for early concept generation \\
\addlinespace

\textit{Function-based heuristics} & Function mapping~\cite{stone_development_2000, holtta-otto_modular_2005} & Systematic functional clustering and simplification & Interface and joining logic not addressed; lacks quantitative assessment \\

\bottomrule
\end{tabular}}
\label{tab:methods_dfa_comparison}
\end{table}

\subsection{Research gap and contribution}
The literature review shows that, despite many modularisation methods and DFA tools, assembly reasoning is still not embedded in early product development. Modularisation is often framed as managing complexity and variety, yet seldom extends to upstream production efficiency, exposing a gap between operational strategy and design execution.

What remains underdeveloped is a method for production-oriented modular design that positions DFA as a strategic input from the outset, an analytical support for proactive decision-making. Such a method must address system-level `assembly in the large'~\cite{whitney_mechanical_2004} and foster multidisciplinary collaboration. While earlier approaches tackle parts of this, none provide an integrated framework for consistent reasoning across concept teams\cite{monetti_barriers_2024}. This paper addresses the gap by prescribing an MFD extension with additional tools.

The extension includes:~(i)~heuristic rules for evaluating TS and defining module structures with assembly in mind, applied through enhancements to DPM and MIM; (ii)~quantitative metrics for assessing architectural assemblability; and (iii)~a taxonomy for classifying interfaces by type, priority, and complexity. Together, these create an expanded toolkit enabling systematic, proactive assembly reasoning within MFD. This answers calls for quantifiable complexity metrics to balance part count, interface richness, and manufacturing effort~\cite{rodriguez-toro_complexity_2003}.

The method is prescriptive, with specific steps and evaluation tools; generalisable, as it builds on the widely used MFD~\cite{erixon_modular_1998, ericsson_controlling_1999}; and practice-informed, drawing on expert interviews, industrial cases, and literature on modularisation, DFX, and assembly systems.

To operationalise the research goals, the following research questions are proposed.
\begin{enumerate}
\item[RQ1] \textbf{How can DFA principles be analytically integrated into the early stages of MFD to support assembly-oriented modular product architectures?}
\newline
Defines the core design challenge:~linking DFA reasoning with MFD’s early steps to embed upstream assembly thinking.

\item[RQ2] \textbf{What heuristics can support the evaluation of TS and module alternatives with respect to DFA in the DPM and MIM?}
\newline
Focuses on developing embedded criteria to guide modular concept evaluation.

\item[RQ3] \textbf{What metrics can assess the assembly-related performance of modular product architectures at system level?}
\newline
Addresses the quantification of assemblability for structured comparison of alternatives.

\item[RQ4] \textbf{How can module interfaces be classified to support strategic DFA decision-making?}
\newline
Targets systematic interface reasoning for robust, automatable modular architectures.
\end{enumerate}

\section{Methodology} \label{sec:methodology}
This study aims to develop a prescriptive, assembly-oriented design suppot to modular product architecture~\cite{blessing_drm_2009}. The method builds on earlier works:~(i)~a systematic review of modularisation, DFA, and product architecture~\cite{monetti_towards_2023}; (ii)~an investigation on industrial challenges of DFA integration~\cite{monetti_barriers_2024}; and (iii)~a pilot implementation of DFA-enhanced MFD~\cite{monetti_evaluating_2025}.

These studies revealed functional, procedural, and analytical gaps in modular design. The proposed method was developed as a design artefact to address them, using a constructive research approach aligned with design science. Insights came from triangulation, and multidisciplinary research group critique. This enabled a balanced view of theoretical needs and implementation constraints, guiding refinement of the method’s structure, logic, and usability.

\subsection{Method development}
The method was developed around three design goals:~(i)~coherence with MFD to ensure compatibility with established practice and terminology; (ii)~integration of assembly logic to support early evaluation of architectural alternatives; and (iii)~stronger decision support for concept selection, module structuring, and interface planning. These address key limitations of existing modularisation and DFA methods, particularly their weak integration at the architectural level.

\begin{figure}
    \centering
    \includegraphics[width=\linewidth]{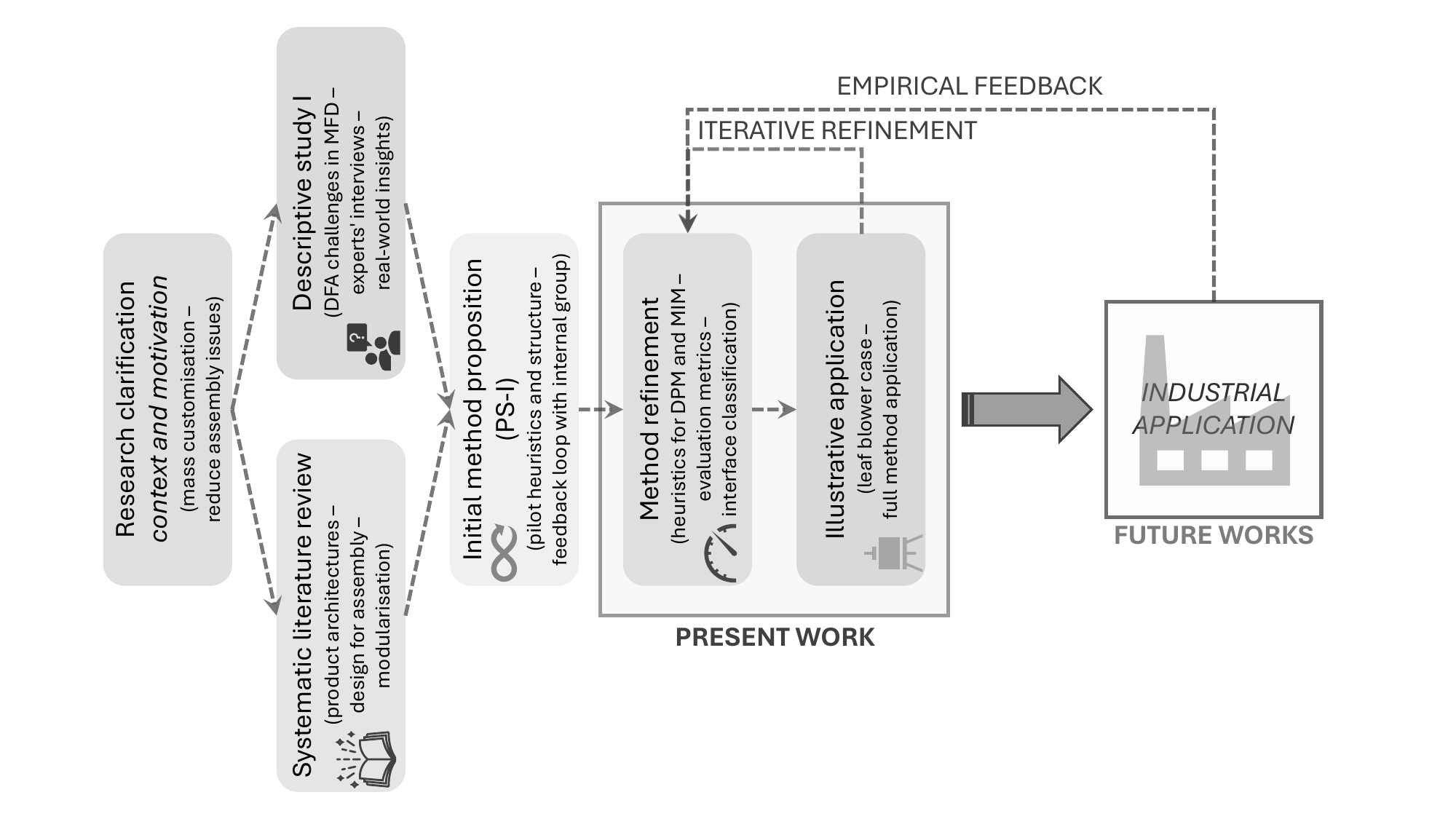}
    \caption{\textbf{Development of the DFA-enhanced MFD method under DRM.} It was developed through systematic review (SLR), industrial insight (DS-I), method proposition (PS-I), and iterative refinement. Arrows indicate feedback loops between stages. Future steps target industrial implementation.}
    \label{fig:method_workflow}
\end{figure}

Figure~\ref{fig:method_workflow} outlines the development trajectory. It began with requirement formulation from literature and industrial interviews, followed by a first set of heuristics and structural elements tested internally. Iterative refinement cycles then validated each component for consistency, feasibility, and alignment with modularisation objectives. The result is a coherent, prescriptive extension of MFD that retains its matrix-based structure while enabling systematic assembly reasoning. Section~\ref{sec:mfd} details the method’s components.

\subsection{Scope of application and demonstration}
The method supports both greenfield (new development) and brownfield (architecture refinement) scenarios. In greenfield contexts, it embeds assembly logic early, guiding function allocation and module boundary definition. In brownfield settings, it diagnoses and reconfigures existing architectures for improved assemblability and strategic alignment. Although the case study here concerns refinement, the method is generalisable to early-stage concept development.

To show coherence and usability, the method was applied to a modular handheld leaf blower, chosen for its clear functional decomposition, interfaces, and mechanical complexity. Its prior use in education enabled controlled experimentation and feedback~\cite{monetti_evaluating_2025}. The aim was not industrial validation but to show how the method supports structured reasoning and trade-offs with assembly in focus.

Section~\ref{sec:casestudy} presents the case, comparing a legacy product architecture with an improved configuration. The analysis assesses method consistency, step completeness, and traceability of assembly-related decisions.

The implementation followed five stages:~(i)~baseline analysis of the existing architecture, including functions and interfaces; (ii)~preliminary improvement assessment by junior engineers; (iii)~application of the proposed method by experienced researchers to explore alternative configurations; (iv)~use of DFA heuristics, metrics, and interface classification to guide generation and comparison of architectures; and (v)~evaluation of both configurations with the proposed metrics to assess assembly efficiency and robustness. While not full industrial validation, the case demonstrates method usability and coherence.
\section{Development of an assembly-oriented expanded MFD method} \label{sec:mfd}
This section outlines the core contributions enabling assembly-oriented architectural reasoning. Each addresses a limitation of traditional modularisation by embedding assembly logic early in development. The contributions expand the analytical scope of MFD~\cite{erixon_modular_1998} (Figure~\ref{fig:mfd_original}) without altering its structure, supporting both concept evaluation and module-level planning.

\begin{figure}
    \centering
    \includegraphics[width=\linewidth]{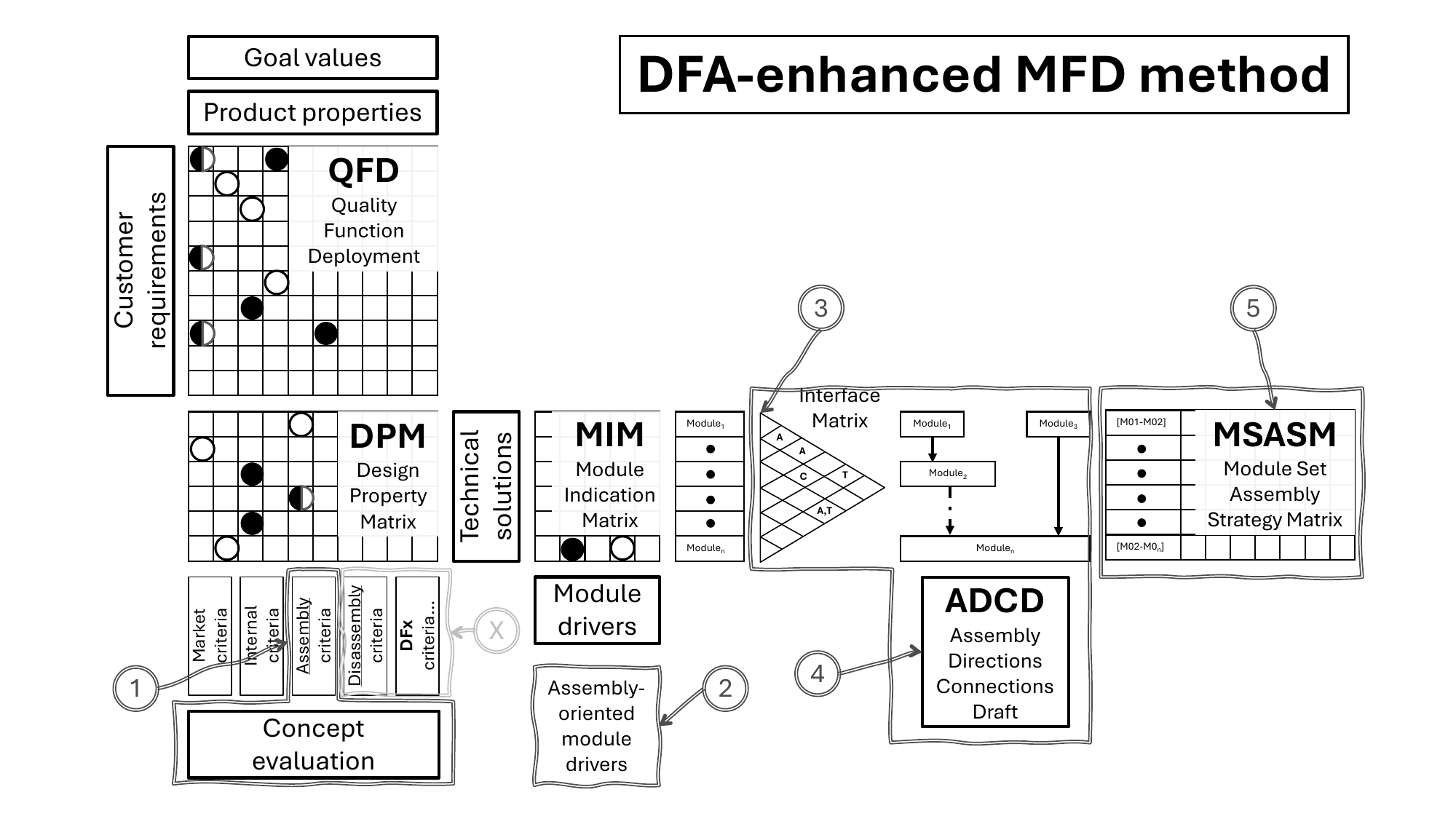}
    \caption{\textbf{Expanded MFD method with DFA enhancements.} Additions include: (1) assembly internal criteria for technical solution evaluation; (2) assembly-oriented module drivers for module formation; (3) a coded interface taxonomy; (4) the ADCD tool for directional interface planning; and (5) the MSASM for quantitative evaluation of architecture-level assemblability. Fully integrated into the MFD workflow, these additions support proactive assembly-oriented decision-making. The concept evaluation stage also remains open for extensions addressing disassembly or other life-cycle criteria (X). \label{fig:mfd_expanded}}
\end{figure}

As shown in Figure~\ref{fig:mfd_expanded}, the enhancements span key workflow phases: concept evaluation/module formation, interface development, and architectural evaluation. In each, new tools, criteria, or classification schemes enable earlier, systematic consideration of assembly before geometry, interfaces, or manufacturing constraints are fixed.

Table~\ref{tab:method_summary} summarises these additions, operationalised through assembly-oriented heuristics and module drivers, an expanded interface taxonomy, tools for interface strategy, and metrics for assessing assemblability and automation readiness.

The following subsections detail each component with implementation illustrated in Section~\ref{sec:casestudy} through the leaf blower case.

\begin{table}
\tbl{Summary of proposed method enhancements: assembly-related design challenges addressed and tools introduced.}
{\begin{tabular}{p{5.5cm} p{9.5cm}}
\toprule
\textbf{Design focus} & \textbf{Method enhancement} \\
\midrule
\textit{Early evaluation of TS and module candidates through assembly-oriented heuristics} & 
1. Assembly-oriented Internal Criteria for design concept evaluation and selection within the DPM \newline
2. Module structuring using assembly-oriented Module Drivers within the DPM \\
\addlinespace
\textit{Interface classification, coding, and scoring} & 
3. Taxonomy and analysis for prioritisation of interfaces \newline
4. Tool for directional interface analysis and assembly planning \\
\addlinespace
\textit{Quantitative evaluation of architecture and interface assemblability} & 
5. Add-on metrics and tool for module-level evaluation of reconfigurability and automation readiness \\
\bottomrule
\end{tabular}}
\label{tab:method_summary}
\end{table}

\subsection{Concept evaluation: integration of DFA criteria after the DPM}
In standard MFD, once Product Properties are linked to functions and TS via the DPM, concept evaluation compares alternatives using Market and Internal Criteria. However, Internal Criteria are often ad-hoc, varying by project, and seldom address production or assembly.

To close this shortcoming, we introduce a predefined set of assembly-oriented criteria into the DPM’s internal assessment. Grounded in established DFA principles, these criteria guide qualitative comparison of TS with respect to scalability, automation readiness, handling complexity, joining direction, and tool needs.

Table~\ref{tab:dfa_criteria} lists nine criteria, each with a description and guiding question to support structured reasoning during early concept selection. They cover part count reduction, insertion ease, fastener standardisation, tool accessibility, and test integration, enabling teams to consider assembly implications before geometry or planning is fixed.

The criteria are selective and flexible. Not all apply to every function or product, and emphasis should align with company strategy. Following Treacy and Wiersema’s value disciplines~\cite{treacy_customer_1993}: firms prioritising Operational Excellence may emphasise complexity reduction and automation readiness (criteria II–IV); Product Leadership may stress standardisation, commonality, and testability (criteria VI–VIII); Customer Intimacy may value upgradability, customisation, and flexibility (criteria I, V, IX). This ensures evaluation supports both feasibility and strategic coherence. The set can be refined for specific strategies.

Evaluation may use either a qualitative Pugh matrix, encouraging relative comparison and cross-functional dialogue, or a 1–5 numerical scale, enabling aggregation and weighting. Both rely on expert judgement, anchored by checklist prompts for consistency and traceability. The assessments are qualitative by design, fitting the early concept stage with limited geometry data, and do not replace part-level DFA but bring its logic upstream.

Results can be incorporated directly into the DPM or as a parallel prioritisation layer. When assembly performance is critical, this structured evaluation clarifies trade-offs and builds consensus early. An extended version of the criteria table, with examples and design risks, is available to practitioners but omitted here for brevity.

Outputs from this evaluation feed into the MIM, where selected TS are grouped into modules. To support this transition, the method introduces assembly-oriented MD, described next.

\begin{table}
\tbl{Assembly-oriented evaluation criteria for early-stage concept selection and qualitative assessment of alternative TS.}
{\begin{tabular}{>{\raggedright\arraybackslash}p{0.035\linewidth}
>{\raggedright\arraybackslash}p{0.15\linewidth}
>{\raggedright\arraybackslash}p{0.30\linewidth}
>{\raggedright\arraybackslash}p{0.30\linewidth}
>{\raggedright\arraybackslash}p{0.115\linewidth}}
\toprule
 & \textbf{Evaluation criteria} & \textbf{Description} & \textbf{Design cues} & \textbf{References} \\
\midrule
I. &
{Scalability opportunities} &
{Supports variant creation with minimal assembly rework} &
{\begin{tabular}[c]{@{}l@{}} • Interfaces allow upgrades \\
• Late integration of options \\ • Future upgrades compatible\end{tabular}}
& \cite{boothroyd_product_2010,pahl_engineering_2007,baldwin_design_2000,ericsson_controlling_1999} \\
\addlinespace

II. &
{Reduce assembly cost/complexity} &
{Enables lower direct labour or test costs through part simplification} &
{\begin{tabular}[c]{@{}l@{}} • Fewer or combined parts\\ • Standard connectors\\ • Minimal tool changes\\ • Safer handling\end{tabular}}
&
\cite{boothroyd_product_2010,eskilander_design_2001,huang_power_2005}\\
\addlinespace

III. &
{Ease of assembly} &
{Enables intuitive, tool-free, and ergonomic assembly} &
{\begin{tabular}[c]{@{}l@{}} • Self-locating \\
• Minimal reorientation \\
• Low insertion force\end{tabular}}
&
\cite{boothroyd_product_2010,van_nes_product_2006,whitney_mechanical_2004,monden_toyota_1998,wada_evolution_2020}\\
\addlinespace

IV. &
{Assembly automation opportunities} &
{Compatible with robotic or automated assembly} &
{\begin{tabular}[c]{@{}l@{}} • Symmetrical geometry \\
• Feeding not orientation-dependent \\
• Suited for grippers \end{tabular}}
&
\cite{boothroyd_product_2010,groover_automation_2016,furtado_dtw_2014}\\
\addlinespace

V. &
{Late point of configuration} &
{Allows for postponing customisation after main assembly} &
{\begin{tabular}[c]{@{}l@{}} • Customisable after assembly\\ • Late-stage feature addition\end{tabular}}
&
\cite{pahl_engineering_2007,fixson_modularity_2007,pine_mass_1993,tseng_novel_2006, gandhi_how_2013}\\
\addlinespace

VI. &
{Limit production investment} &
{Avoids dedicated tools or fixtures for one solution} &
{\begin{tabular}[c]{@{}l@{}} • Common fixtures\\ • Standardised equipment\\ • General-purpose tools\end{tabular}}
&
\cite{boothroyd_product_2010,ulrich_product_2020}\\
\addlinespace

VII. &
{Limit production test cost} &
{Reduces need for complex, costly test setups} &
{\begin{tabular}[c]{@{}l@{}} • Allows in-line testing\\ • Avoid recalibration\end{tabular}}
&
\cite{eskilander_design_2001,bralla_design_1999,liker_toyota_2004}\\
\addlinespace

VIII. &
{Improved yield} &
{Reduces defect risk and rework} &
{\begin{tabular}[c]{@{}l@{}} • Poka-yoke elements \\
• Pre-testable \\
• Reduced tolerance sensitivity\end{tabular}}
&
\cite{shingo_zero_2007,boothroyd_product_2010}\\
\addlinespace

IX. &
{Assembly line balance} &
{Enables balanced workload and avoids bottlenecks} &
{\begin{tabular}[c]{@{}l@{}} • Even task time \\
• Sequencing flexibility\end{tabular}}
&
\cite{boothroyd_product_2010,pahl_engineering_2007,wada_evolution_2020}\\

\bottomrule
\end{tabular}}
\label{tab:dfa_criteria}
\end{table}

\subsection{Assembly-oriented module drivers for the MIM}
After TS evaluation, module candidates are formed using MDs: criteria reflecting strategic, operational, or technical reasons for grouping or separating solutions.

Traditionally, MDs emphasise lifecycle, customisation, or sourcing. These remain essential, but production-oriented concerns are often under-represented. Table~\ref{tab:module_drivers} consolidates established MDs by value chain steps and introduces a dedicated set of assembly-oriented drivers.

The new drivers draw on DFA literature~\cite{boothroyd_product_2010,eskilander_design_2001}, lean principles~\cite{womack_lean_1997}, and related work~\cite{mesa_development_2018}. They complement rather than replace existing MDs, ensuring architectures support automation-ready production strategies.

Application follows standard MFD logic:~TS pairings are checked for synergies or conflicts across MDs. For instance, common feeding orientation or fastening logic may support grouping (via Integrated fastening or Handling), while differing joining directions or takt-time effects may justify separation (via Assembly line balancing or Pre-assembly).

Assembly-oriented MDs act as prompts, not fixed rules. Prioritisation should reflect company strategy: automation-focused firms may emphasise Interface standardisation or Integrated fastening; labour-intensive contexts may favour Assembly sequence or Pre-assembly.

Trade-offs between drivers are expected and should be made explicit. Pre-assembly, for example, may conflict with Separate testability; Interface standardisation may constrain customisation. These tensions are essential inputs to architectural decision-making.

The balance between strategic, product-centred and operational, production-focused MDs should be shaped by organisational priorities. Firms aiming for platform scalability may stress lifecycle MDs, while those targeting automation or takt control may rely more on assembly-oriented drivers. Enriching the MIM with these drivers makes it both diagnostic and generative, helping teams design architectures that are functionally coherent and executable in production.

\begin{table}
\tbl{Overview of module drivers aligned with value chain steps (adapted from~\cite{andersen_module_2022}), with descriptions to guide grouping or separation of TS.}
{\begin{tabular}{
>{\raggedright\arraybackslash}p{0.14\linewidth}
>{\raggedright\arraybackslash}p{0.20\linewidth}
>{\raggedright\arraybackslash}p{0.46\linewidth}
>{\raggedright\arraybackslash}p{0.10\linewidth}}
\toprule

\textbf{Value chain step} & \textbf{Module driver} & \textbf{Description} & \textbf{Reference} \\
\midrule
\multicolumn{4}{l}{\textit{Established MDs in literature}} \\
\midrule

\multirow{5}{=}{Product development} 
& Carry over & Group TS reused from previous generations to reduce development time, cost, and risk. & \cite{erixon_modular_1998,kreng_qfd-based_2004,politze_function_2012,andersen_module_2022} \\
& Planned development & Separate TS expected to evolve independently, enabling easier updates and technology integration. & \cite{erixon_modular_1998, andersen_module_2022} \\
& Technology push & Group TS involving emerging technologies to isolate uncertainty and accelerate innovation. & \cite{erixon_modular_1998, andersen_module_2022} \\
& Technical specification & Group TS with similar specification or performance targets to streamline verification and variation handling. & \cite{erixon_modular_1998,andersen_module_2022} \\
& Styling/customisation & Separate TS subject to customisation or aesthetic variation to localise complexity and support market differentiation. & \cite{erixon_modular_1998,kreng_qfd-based_2004,andersen_module_2022} \\

\midrule
\multirow{5}{=}{Manufacturing}
& Common process & Group TS requiring the same manufacturing processes (e.g., casting, welding) to simplify production flow. & \cite{erixon_modular_1998,kreng_qfd-based_2004, andersen_module_2022} \\
& Common unit & Group TS reused across products or variants to maximise reuse and economies of scale. & \cite{erixon_modular_1998,andersen_module_2022,kreng_qfd-based_2004} \\
& Automation & Group TS that support the same automation level; separate those requiring distinct manual operations. & \cite{boothroyd_product_2010, blees_development_2008} \\
& Late differentiation & Group or separate TS to enable delayed integration of variant-defining components. & \cite{blackenfelt_managing_2001} \\
& Special process requirements & Group TS requiring unique production conditions (e.g., cleanroom, hazardous materials) to ensure safety and quality. & \cite{blees_development_2008} \\

\midrule
\multirow{2}{=}{Functionality}
& Separate testability & Isolate TS requiring individual testing prior to integration to improve quality control and fault detection. & \cite{erixon_modular_1998, lange_modular_2014, blackenfelt_managing_2001} \\
& User perception & Group or separate TS that are perceived together or co-activated during use to support intuitive operation or customisation. & \cite{politze_function_2012} \\

\midrule
\multirow{3}{=}{Logistics}
& Strategic supplier & Group TS sourced from the same supplier to simplify coordination and contractual management. & \cite{andersen_module_2022} \\
& Purchasing & Group or separate TS based on sourcing or outsourcing decisions to align with procurement strategy. & \cite{kreng_qfd-based_2004} \\
& Storage & Separate TS with special storage or preservation needs (e.g., humidity control, shelf life). & \cite{blees_development_2008} \\

\midrule
\multirow{3}{=}{After sales}
& Serviceability & Separate TS expected to undergo maintenance or replacement to enable easy access and minimal disruption. & \cite{erixon_modular_1998, lange_modular_2014} \\
& Upgrading/flexibility & Group or separate TS with potential for future performance upgrades or functional rearrangements. & \cite{erixon_modular_1998,kreng_qfd-based_2004} \\
& Recycling & Separate TS to facilitate end-of-life disassembly, material sorting, and recovery. & \cite{kreng_qfd-based_2004,andersen_module_2022} \\ 

\midrule
\multicolumn{4}{l}{\textit{Assembly-oriented drivers, proposed expansion}} \\
\midrule
\multirow{6}{=}{Assembly}
& Assembly sequence & Group TS to simplify joining order, reduce interdependencies, and support directional assembly. & \cite{boothroyd_product_2010, womack_lean_1997, eskilander_design_2001} \\

& Assembly line balancing & Group TS to distribute tasks evenly across stations and avoid bottlenecks. & \cite{boothroyd_product_2010, womack_lean_1997} \\

& Pre-assembly & Cluster TS that can be joined offline and later integrated as a subassembly. & \cite{boothroyd_product_2010} \\

& Integrated fastening & Group TS using shared fasteners or tools to reduce joining variation and tool changes. & \cite{boothroyd_product_2010} \\

& Interface standardisation & Group TS around common interfaces to support repeatable and automatable connections. & \cite{boothroyd_product_2010} \\

& Handling (complements automation) & Group TS with similar handling needs (e.g., gripping, orientation) to ease feeding and reduce changeover. & \cite{boothroyd_product_2010,mesa_development_2018} \\

\bottomrule
\end{tabular}}
\label{tab:module_drivers}
\end{table}

\subsection{Interface taxonomy: types and priorities}
Interfaces between modules play a central role in determining assembly effort, modular flexibility, and architectural robustness. In modular systems, interfaces connect parts and they define how loads, energy, and signals are transferred, influencing joining methods, sequencing, and production. Despite their importance, interface reasoning is underdeveloped in early architectural phases, possibly leading to costly redesigns during industrialisation.

This subsection introduces a unified interface taxonomy grounded in three complementary sources:~the attachment/transfer/control model used in industrial modularisation practice; the assembly-centric interface types by Favi and Germani~\cite{favi_method_2012}; and the complexity-based interface sensitivity proposed by Mesa et al.~\cite{mesa_development_2018}.

The scope is deliberately limited to interfaces relevant to assembly operations. Spatial relationships between modules (e.g., for thermal, acoustic, or maintenance considerations) are excluded, as they do not involve joining or signal transfer. These are better modelled as Product Properties with defined Goal Values (e.g., $Motor~volume < 0.5~L$), or handled within CAD and configuration environments.

The taxonomy groups interfaces into five main types, detailed in Table~\ref{tab:interface_scores}. These include:~mechanical transmission (M1) and structural mechanical (M2) interfaces, which often dictate early sequencing due to their load-bearing role; magnetic (B) and optical (O) interfaces, which are sensitive to alignment and require controlled assembly conditions; standard electrical interfaces (E), typically lower in planning priority; and control interfaces (C), which convey sensor or actuator signals and user commands. The latter demand careful routing and testing, thus requiring intermediate priority in assembly planning. These categories enable structured reasoning about joining needs before detailed geometry is defined.

Each interface type is given a priority score (1–4) reflecting its influence on sequencing, access, and quality assurance, independent of functional importance. Following Favi and Germani~\cite{favi_method_2012}, these priorities guide architecture-level sequencing in a layered shell model:~high-priority mechanical interfaces are assembled first to create stability, followed by lower-priority connections such as control or electrical. This logic is later applied in the IM and ADCD, where priorities shape spatial layout and joining order.

The taxonomy is not intended to replace geometric models but to function as a reasoning overlay during early design phases. When used alongside the DPM and MIM, it enhances architecture-level decisions and supports the development of production-aligned, automation-friendly modular product structures.

While the taxonomy is designed for use in early concept development, it does assume a basic level of interface definition. In greenfield projects, this typically relies on rough sketches, functional block diagrams, or design heuristics that indicate likely joining principles, transfer mechanisms, and component characteristics (e.g., size, mass, number of connectors). Absolute accuracy is not required, but some level of embodiment is likely necessary to apply the coding logic meaningfully. As more detailed design data becomes available, interface codes can be updated to reflect refined requirements. This enables iterative reassessment of architectural alternatives without requiring full CAD definition.

\begin{table}
\tbl{Unified interface taxonomy with priority values and complexity scoring factors. The upper section lists interface types. The lower section defines task-level assembly complexity factors.}
{\begin{tabular}{p{0.20\linewidth} p{0.08\linewidth} p{0.55\linewidth} p{0.10\linewidth}}
\toprule
\textbf{Type} & \textbf{Code} & \textbf{Description} & \textbf{Priority\textsuperscript{1}} \\
\midrule
Mechanical transmission & M1 & Transfers motion or torque (e.g., shafts, gears) & 3 \\
Structural mechanical & M2 & Static structural joining (e.g., housings, brackets) & 2 \\
Magnetic/optical & B, O & High-precision interfaces sensitive to alignment (e.g., magnets) & 4 \\
Electrical & E & Power supply and passive energy flow (e.g., heating elements) & 1 \\
Control & C & Signal and command interfaces or HMIs (e.g., sensors, actuators) & 3 \\
\midrule
\textbf{Factor} & \textbf{Code} & \textbf{Description} & \textbf{Score\textsuperscript{2}} \\
\midrule
\textbf{Handling} & & \textit{Number of hands required} & \\
& H1 & One hand & 0.25 \\
& H2 & Two hands & 0.50 \\
& H3 & More than two hands & 1.00 \\
\textbf{Alignment} & & \textit{Precision needed for successful joining} & \\
& A0 & None & 0.00 \\
& A1 & Low – ½ rotation or simple orientation & 0.25 \\
& A2 & Medium – ¾ rotation or moderate guidance & 0.50 \\
& A3 & High – full rotation or exact fit & 0.75 \\
& A4 & Very high – multi-axis or compound fit & 1.00 \\
\textbf{Tooling need} & & \textit{Type of tool required for joining} & \\
& T0 & None & 0.00 \\
& T1 & Conventional tool (e.g., screwdriver) & 0.50 \\
& T2 & Specialised or powered tool & 1.00 \\
\textbf{Interface fit ease} & & \textit{Adjustment constraint or guidance level} & \\
& I0 & Intuitive/self-aligning & 0.25 \\
& I1 & Guided or constrained fit & 0.50 \\
& I2 & Manual or force fit (e.g., friction) & 1.00 \\
\textbf{Fixing device need} & & \textit{Requirement for additional fixation} & \\
& D0 & No fixation required & 0.00 \\
& D1 & Standard fasteners (e.g., clips, screws) & 0.50 \\
& D2 & Specialised fixtures or alignment tools & 1.00 \\
\bottomrule
\end{tabular}}
\tabnote{\textsuperscript{1} Interface priority reflects assembly planning sensitivity, not functional criticality. High-priority interfaces require earlier sequencing
or stricter quality control in assembly planning. \newline
\textsuperscript{2} Scores are adapted from~\cite{mesa_development_2018} and used in the alpha-numeric coding for assembly complexity evaluation.}
\label{tab:interface_scores}
\end{table}

\subsection{Interface taxonomy: complexity and coding scheme} \label{sec:taxcompl}
To support reasoning about inter-module connections, this study introduces a coding scheme capturing key assembly-related interface characteristics. Building on the earlier taxonomy (mechanical, electrical, optical, magnetic, control), it integrates complexity factors from Mesa et al.~\cite{mesa_development_2018}, which quantify dis/assembly effort in open modular architectures. The resulting alphanumeric code expresses assembly sensitivity and helps prioritise interface redesign or sequencing without detailed geometry.

Table~\ref{tab:interface_scores} summarises the coding logic, defining discrete levels for five parameters, each with associated scores. These indicators enable calculation of overall interface complexity.

Each interface is represented by a structured alphanumeric string encoding its type, assembly priority, and task-level attributes:
\begin{equation}
\mathtt{{InterfaceType}-P_{value}-H_{value}-A_{value}-T_{value}-I_{value}-D_{value}}.
\end{equation}
HHere, \texttt{InterfaceType} denotes the interface category (e.g., \texttt{M1} for motion-transfer mechanical, \texttt{O} for optical). The priority value ($P_{value}$, 1–4) indicates strategic importance in assembly planning. Handling ($H_{value}$) specifies required hands: H1 (one), H2 (two), H3 (more than two). Alignment ($A_{value}$) captures precision needs, from 0 (none) to 4 (multi-axis/high). Tooling ($T_{value}$) reflects tool use: 0 (none), 1 (standard), 2 (specialised). Interface fit ($I_{value}$) measures connection ease: 0 (self-aligning), 1 (normal), 2 (constrained/complex). Fixing device ($D_{value}$) indicates fastening complexity: 0 (none), 1 (standard), 2 (specialised).

For example, a \texttt{C} (control) interface with priority 3, two-hand handling (H2), moderate alignment (A2), standard tool (T1), intuitive fit (I0), and no fastener (D0) is coded as \texttt{C-P3-H2-A2-T1-I0-D0}.
A \texttt{M1} (mechanical with motion) interface with priority 3, one-hand handling (H1), low alignment (A1), specialised tool (T2), intuitive fit (I0), and standard fastener (D1) is coded as \texttt{M1-P3-H1-A1-T2-I0-D1}.

\begin{figure}
    \centering
    \includegraphics[width=\linewidth]{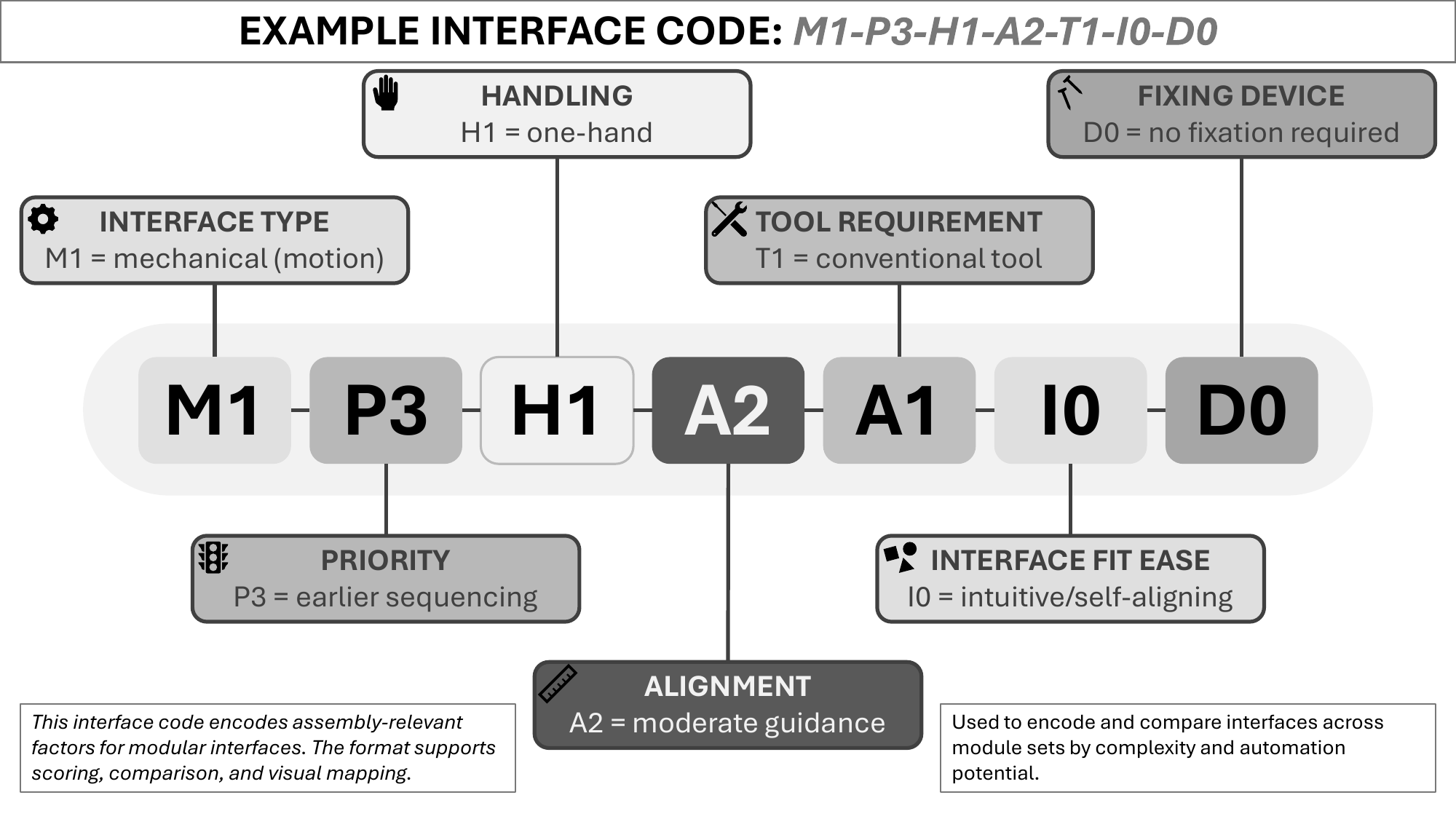}
    \caption{\textbf{Cheat sheet for interpreting the interface code string (e.g., \texttt{M1-P3-H1-A2-T1-I0-D0})}. Each component encodes an assembly-relevant attribute:~interface type, priority, handling, alignment, tooling, fit, and fixing. The structure enables systematic comparison and complexity scoring. \label{fig:interface_code_cheatsheet}}
\end{figure}

This coding system offers a structured way to assess interface complexity during architecture synthesis. Figure~\ref{fig:interface_code_cheatsheet} illustrates the code format and its mapping to assembly effort. By unifying priority and complexity dimensions in a single string, the method supports cross-functional dialogue and structured decision-making in early modular development.

The interface complexity score quantifies assembly effort by summing the weighted values of five attributes. While abstracting from geometry, it provides a useful proxy for assessing assembly sensitivity in early architectural planning. \\
\textbf{Example 1:} Mechanical interface (\texttt{M1}), priority 3; one-handed handling ($H1 = 0.25$), medium alignment ($A2 = 0.5$), standard tool ($T1 = 0.5$), intuitive fit ($I0 = 0.25$), no fixing device ($D0 = 0$).
Code: \texttt{M1-P3-H1-A2-T1-I0-D0}
Score: $0.25 + 0.5 + 0.5 + 0.25 + 0 = \textbf{1.5}$. \\
\textbf{Example 2:} Optical interface (\texttt{O}), priority 4; two-handed handling ($H2 = 0.5$), high alignment ($A3 = 0.75$), specialised tool ($T2 = 1.0$), constrained fit ($I2 = 1.0$), standard fastener ($D1 = 0.5$).
Code: \texttt{O-P4-H2-A3-T2-I2-D1}
Score: $0.5 + 0.75 + 1.0 + 1.0 + 0.5 = \textbf{3.75}$.

These scores feed into downstream assessment tools that evaluate modular architectures based on cumulative interface effort, joining directionality, and integration complexity.

\subsection{Application of the taxonomy in the Interface Matrix}
The IM is a key tool for documenting and analysing inter-module connections. Incorporating the unified taxonomy and coding scheme extends it into a structured evaluator of interface complexity.

Each module pair is classified with the taxonomy and assigned a complete code, ensuring consistency and enabling comparison across architectures. The IM can be rearranged to visualise connection density and criticality: modules with many links are placed centrally as base or structural units, while high-priority or complex interfaces (e.g., motion-transfer mechanical or high-precision couplings) inform sequencing. This supports macro-level assembly strategies such as the base-unit strategy (centre-out from a structural hub), hamburger strategy (vertical stacking from complex to simpler modules), and bus strategy (parallel sub-assemblies feeding a central module).

Interface complexity is calculated as the sum of the five task-level scores (H, A, T, I, D). These values feed into system-level metrics (introduced later) that guide iteration by quantifying assembly burden, interface resilience, and variant flexibility.

The enhanced IM thus supports clustering via low-complexity interfaces, boundary adjustment where complexity accumulates, and informed layout planning. It enables a strategy-clean modularisation process, where architectural decisions reflect not only functional and market needs but also production-specific goals and constraints.

\subsection{Assembly Directions and Connections Draft}
The Assembly Directions and Connections Draft (ADCD) is a structured visual reasoning tool developed to support early-stage layout planning of modular product architectures. It represents directional dependencies, interface characteristics, and strategic assembly constraints before physical embodiment is fixed.

Unlike generic schematics or exploded views, the ADCD builds upon the interface coding scheme introduced earlier. Each connection is represented as a directed edge between modules, annotated with its full code (e.g., \texttt{M1-P3-H1-A2-T1-I0-D0}). Assembly directionality (top-down, lateral, or frontal) is explicitly encoded, supporting ergonomic assessment and sequencing logic.

The ADCD adopts the shell-based stratification model proposed by Favi and Germani~\cite{favi_method_2012}, grouping modules into layers that reflect their role in the product structure. Inner shells contain base modules that receive insertions (e.g., structural frames), while outer shells host peripheral, reconfigurable, or late-inserted components (e.g., nozzles, covers). The stratification supports variant management and early sequencing alignment. From a CE perspective, placing Product Leadership and Customer Intimacy modules in outer shells facilitates selective disassembly, upgrade, or reuse, contributing to lifecycle modularity and sustainability. These logics can also structure intended sequences and highlight ergonomic or tool-access conflicts.

Special constraints such as simultaneous insertion, limited visibility, or use of torque-controlled tools can be flagged directly within the ADCD. This reveals interface-related complexity early, allowing mitigation through redesign, module reassignment, or fixturing. While typically visualised as schematics or 3D diagrams, the ADCD is implementation-agnostic and can also be expressed through enriched node–edge graphs containing coded metadata.

Overall, the ADCD enables cross-functional teams to reason collectively about spatial layout, interface feasibility, and assembly sequencing. It strengthens traceability between interface-level decisions and physical configuration, supporting more coherent design iterations and alignment with ergonomic and automation goals.

\subsection{Module Set Assembly Strategy Matrix} \label{sec:msasm}
The Module Set Assembly Strategy Matrix (MSASM) is introduced as a complementary design reasoning tool that supports early evaluation of assembly feasibility across alternative module combinations. It enables proactive assessment of manufacturability, ergonomics, and automation readiness prior to geometric embodiment or process definition.

While traditional MFD emphasises customer value and functional fulfilment, the MSASM brings production strategy explicitly into the process. It shifts from late-stage validation to proactive architectural planning, ensuring that design decisions are aligned with long-term manufacturing goals.

The MSASM is structured as a weighted scoring matrix, evaluating a pair or group of modules that require joining (here defined as a \textit{module set}, e.g., \texttt{M01–M02}) against a defined set of assembly-oriented criteria. These include strategic aspects such as automation potential, handling efficiency, joining directionality, and interface resilience.

A module set is conceptually distinct from an interface. Whereas an interface refers to a single connection such as a mechanical joint or electrical contact, a module set denotes a group of modules considered together for evaluation. This grouping may include multiple interfaces (e.g., M2 and C) and reflects operational considerations such as simultaneous insertion, or spatial alignment constraints.

Although the MSASM often evaluates pairs of modules, this is not a limitation:~the tool can be applied to triads or larger sets when evaluating complex insertions, or cases involving shared interface zones. Thus, it not only supports assessment of interface feasibility, but also enables iterative refinement of module boundaries and sub-assembly strategies.

A standardised version of the matrix is provided. However, it is intended to be adaptable:~designers can modify weightings or exclude criteria based on their strategies or constraints. This allows it to function as a general decision-support and a tailored evaluation framework. By integrating it with the IM and ADCD, teams can reason about modular structures, evaluate trade-offs, and plan cost-effective early design decisions.

\subsubsection{Structure of the MSASM}
Each row in the MSASM corresponds to a module set while columns represent criteria grouped into three strategic domains:~(i)~assembly feasibility, which captures physical and handling complexity based on DFA principles; (ii)~value chain factors, such as transportability and serviceability; and (iii)~automation potential, including characteristics that support high-throughput or robotic assembly. While including a wide range of criteria, the primary factors that dominate assessment outcomes are the number and type of interfaces to be joined; the directionality and complexity of the required assembly motion; and the accessibility for operators or tools during handling, insertion, or service.

Evaluation is performed using a triadic ordinal scale adapted from DFA literature~\cite{boothroyd_product_2010,eskilander_design_2001}, where each criterion is rated as 9 (favourable), 3 (moderate), or 1 (unfavourable). A score of 9 denotes low complexity or high efficiency, while 1 signals significant assembly challenges. To reflect strategic priorities, each criterion is assigned a weight from 1 (low importance) to 5 (very high importance). The overall module set score is computed as a weighted average:
\begin{equation} \label{eq:msasm_score}
    MSASM_{\text{score}} = \frac{\sum_{i=1}^{n}{w_i \cdot s_i}}{\sum_{i=1}^{n}{w_i}},
\end{equation}
where $s_i$ is the score and $w_i$ the weight for criterion $i$. This formulation supports consistent yet customisable assessments aligned with product strategy.

The full set of criteria is presented in Table~\ref{tab:msasm_criteria}. Module sets are typically identified during or after the MIM phase. A consistent base module is designated (e.g., the most central unit in the IM) to establish orientation and insertion reference, following the logic of the ADCD.

The MSASM score supports early prioritisation of module sets. Scores above 8.0 indicate assembly readiness; 7.0–7.9 suggest minor refinement may help; 6.0–6.9 highlight elevated complexity; and below 6.0 signals critical feasibility issues. Based on the this, two broad response strategies are likely available:~redesign the product or module set to reduce complexity; or develop special fixtures, tools, or support systems to accommodate the current design. In line with DFA principles, redesign should be prioritised where feasible, as it tends to offer more robust and cost-effective outcomes across production volumes and contexts.

The thresholds are adaptable:~teams can tailor weightings to match strategic priorities: for instance, automation-focused organisations may emphasise criteria such as gripping or insertion complexity, while companies operating in low-volume or manual contexts might prioritise handling and operator compatibility. When used iteratively, it can be used to highlight high-risk module sets, support early sequencing decisions, and enable architecture-level evaluation before physical embodiment is finalised. Combined with the other tools, it provides a robust foundation for strategy-clean modularisation.

Finally, the tool is not limited to final joining operations. It can be applied across the full value chain, including pre-assembled module sets produced by suppliers or in decentralised sub-lines. This enables early decisions on where to locate operations based on complexity, capabilities, or outsourcing strategy supporting both make-or-buy decisions and sequencing plans. By evaluating each module set independently, it allows for task allocation across internal and external resources.

\begin{table}
\tbl{Module Set Assembly Strategy Matrix (MSASM) scoring template.}
{\begin{tabular}{>{\raggedright\arraybackslash}p{0.02\linewidth} >{\raggedright\arraybackslash}p{0.25\linewidth} >{\raggedright\arraybackslash}p{0.2\linewidth} >{\raggedright\arraybackslash}p{0.2\linewidth} >{\raggedright\arraybackslash}p{0.2\linewidth}  >{\raggedleft\arraybackslash}p{0.07\linewidth}}
\toprule
& \textbf{Criterion\textsuperscript{1,2,7}} & \textbf{Score 9 (Favourable)} & \textbf{Score 3 (Moderate)} & \textbf{Score 1 (Unfavourable)} & \textbf{Weight\textsuperscript{3}} \\
\midrule
\multicolumn{6}{l}{\textit{Assembly feasibility}} \\
\midrule
1 & Attachment interface connections & One interface; simple handling and fastening & Two interfaces; moderate complexity & Three or more interfaces; complex alignment or tooling & 4 \\
2a & Assembly direction to base module\textsuperscript{4} & Vertical insertion into fixture; gravity-assisted & Side or angled insertion; minor alignment & Underside insertion; complex fixturing or manual support & 3 \\
2b & Relative direction of subsequent modules\textsuperscript{5} & Same direction as base; gravity-assisted & Same direction as base; not gravity-assisted & Different from base; causes ergonomic or automation issues & 3 \\
3 & Assembly motion and reachability & Straightforward linear motion; fully accessible & Two motions; may require re-orientation & Complex path; limited access; multiple re-orientations & 4 \\
4 & Part orientation & Symmetrical or self-aligning parts; minimal effort & Some asymmetry; manageable with guidance & Requires pre-alignment; difficult to orient & 2 \\
5 & One operator compliant & Handled easily by one person & Requires temporary support or assistance & Needs two operators or difficult solo handling & 5 \\
6 & Insertion & Gravity-assisted or push-fit & Requires manual guidance & Needs tools or force & 3 \\
7 & Handling & Lightweight and easy to grip & Medium weight; requires partial support & Heavy or slippery; needs special tools & 2 \\
\midrule
\multicolumn{6}{l}{\textit{Logistics and value chain}} \\
\midrule
8 & Transport and storage & Compact and stackable & Bulky or awkward & Requires special storage or handling & 3 \\
9 & Disassembly and recycling & Tool-free and non-destructive & Tools required; manageable & Destructive or complex disassembly & 3 \\
10 & Maintenance accessibility & Fully accessible without disassembly & Partial disassembly needed & Not accessible without full teardown & 2 \\
\midrule
\multicolumn{6}{l}{\textit{Automation compatibility}} \\
\midrule
11 & Automatic feeding & Easy: symmetric and rigid & Needs orientation aids or alignment & Flexible, sticky, or asymmetric parts & 3 \\
12 & Gripping ease & Compatible with standard grippers & Requires specialised grippers or surfaces & Multiple gripper changes required & 4 \\
13 & Automatic insertion and fastening & Simple: gravity or snap-fit & Requires moderate alignment & Needs force, tools, or special equipment & 3 \\
\midrule
\multicolumn{2}{l}{\textbf{Score range\textsuperscript{6}}} & \multicolumn{4}{l}{\textbf{Interpretation}} \\
\midrule
\multicolumn{6}{l}{\textit{Suggested thresholds for interpreting weighted average MSASM scores.}} \\
\midrule
\multicolumn{2}{l}{$\geq$ 8.0} & \multicolumn{4}{l}{\textbf{Good}:~suitable for assembly without major redesign. Proceed as planned.} \\
\multicolumn{2}{l}{7.0–7.9} & \multicolumn{4}{l}{\textbf{Acceptable}:~minor improvements recommended to optimise performance or reduce cost.} \\
\multicolumn{2}{l}{6.0–6.9} & \multicolumn{4}{l}{\textbf{Caution}~redesign or mitigation strongly advised to address potential risks.} \\
\multicolumn{2}{l}{$\leq$ 5.9} & \multicolumn{4}{l}{\textbf{Critical}:~critical assembly-issues unless significant adaptation, rework, or fixturing.} \\
\bottomrule
\end{tabular}}
\tabnote{\textsuperscript{1} Each module set is evaluated against 13 criteria grouped into three categories. Each criterion is scored using a 1–3–9 scale, consistent with ordinal evaluation principles in DFA literature~\cite{boothroyd_product_2010,eskilander_design_2001}. \newline
\textsuperscript{2} Criteria 1 to 7 belong to the category \textit{`Assembly feasibility'}, criteria 8 to 10 belong to the category \textit{`Logistics and value chain'}, and criteria 11 to 13 belong to the category \textit{`Automation compatibility'}. \newline
\textsuperscript{3} Design teams assign a weight ranging from 1 (low) to 5 (high) to each criterion depending on the strategic priorities of the company or product line (e.g., favouring automation, reducing operator dependency, improving maintenance). \newline
\textsuperscript{4} Designers should identify the bus module set early during interface planning. This usually corresponds to the core functional platform or most-connected module (based on IM and number of connections). Typically placed in a fixture or base station. When scoring, make sure that all other module sets are evaluated in reference to the orientation of this base module. \newline
\textsuperscript{5} Every additional module added to the system after the base. \newline
\textsuperscript{6} The total score per module set is calculated using the weighted sum (see Eq.~\ref{eq:msasm_score}), supporting comparative assessments and flagging high-complexity configurations for redesign. \newline
\textsuperscript{7} Each criterion also supports broader benefits, namely efficiency, quality, safety and sustainability. These can guide weighting decisions and identify areas where redesign is critical. Full list in the complete tool.}
\label{tab:msasm_criteria}
\end{table}

\subsubsection{Complementary task-level assessment: Total Assembly Complexity (TAC)}
To complement the strategy-level evaluation of the MSASM, this method introduces a task-level metric:~\textbf{Total Assembly Complexity (TAC)}. TAC quantifies the cumulative operational effort required to execute individual operations within each module set.

Each module set is broken down into assembly tasks defined by connections. Tasks are scored using the coding scheme, covering five factors:~handling (H), alignment (A), tool use (T), interface fit (I), fixing requirements (D). Task complexity $C_t$ is calculated as:
\begin{equation} \label{eq:complexity}
    C_t = H + A + T + I + D.
\end{equation}
Complexity scores are then aggregated at two levels:~the module set and the overall architecture.
\begin{align}
    TAC_{\text{module set}} &= \sum_{i=1}^{n} C_{t_i},~\text{and} \\
    TAC_{\text{architecture}} &= \sum_{j=1}^{m} TAC_{\text{module set}_j},
\end{align}
where $n$ is the number of discrete tasks in a given module set, and $m$ is the number of module sets in the overall architecture.

TAC highlights sources of assembly burden overlooked by higher-level metrics. As an unweighted, strategy-neutral measure, it is useful for spotting bottlenecks, task-intensive configurations, and opportunities for simplification, boundary revision, or sequence optimisation. At the architecture level, it enables preliminary quantitative comparison of alternatives.

Together, MSASM aligns choices with production strategy, while TAC exposes task difficulty. This multi-level evaluation ensures early modularisation decisions are both strategically justified and operationally feasible.
\section{Case study: application to a handheld leaf blower} \label{sec:casestudy}
This case study demonstrates the application of the proposed DFA-enhanced MFD workflow to a commercially available handheld electric leaf blower. The objective is not industrial validation, but to illustrate the internal coherence, practical usability, and analytical resolution of the method.

The product represents a typical mid-range garden appliance, characterised by moderate mechanical complexity and partial modularisation in its baseline design. These features, alongside its disassemblability and component clarity, make it well-suited for research and educational use in modular architecture development. Although this is a brownfield study, the workflow can be equally applied to greenfield development; here, an existing product was selected to prioritise coherence and illustrate the method across all steps using a tangible reference.

\begin{figure}
    \centering
    \includegraphics[width=0.7\linewidth]{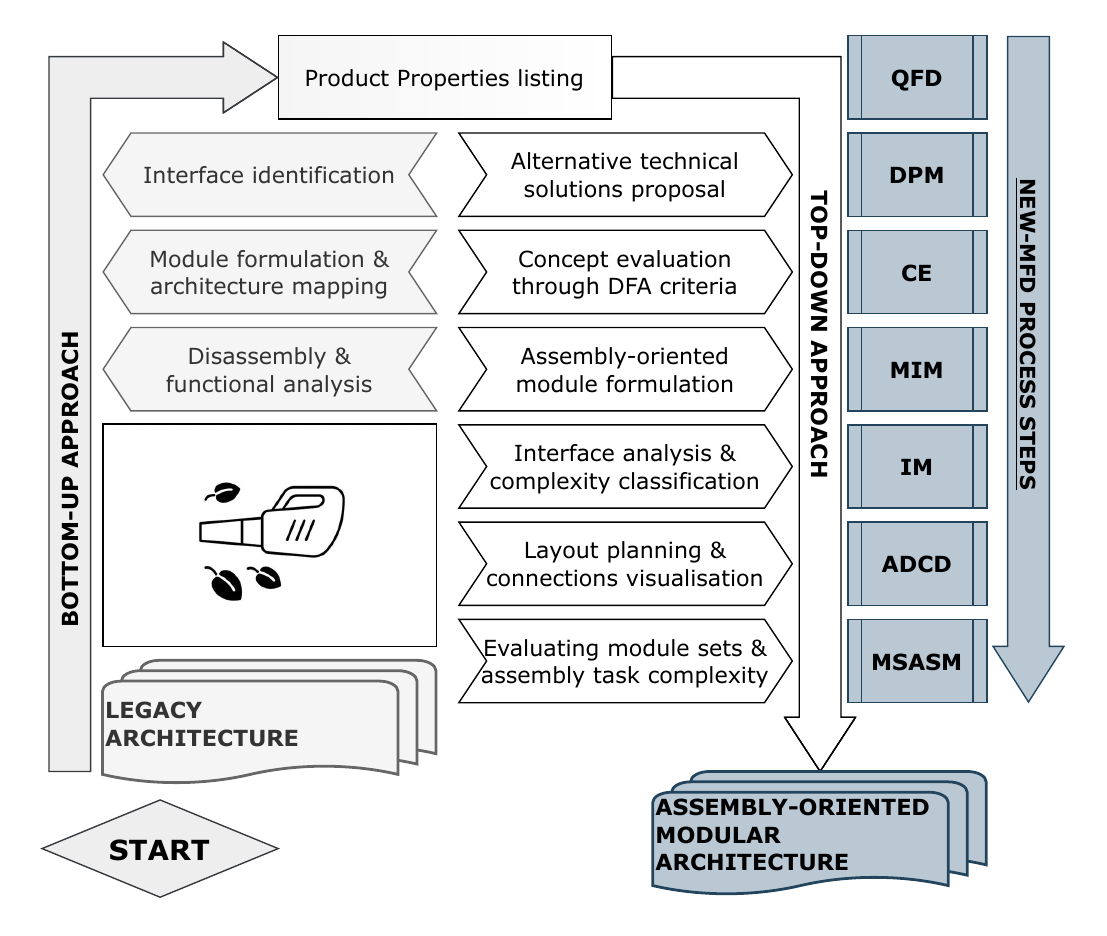}
    \caption{\textbf{Comparison of legacy and expanded modularisation workflows.}
Bottom-up teardown-based modularisation (left) and assembly-oriented top-down architecture development (right). The proposed method incorporates the added tools. \label{fig:case_study_process}}
\end{figure}

Three goals guide the case study, namely to apply the full DFA-enhanced MFD workflow to a real product; to illustrate how each step enables structured reasoning and design evaluation; and to identify potential improvements based on heuristics, interface classification, and modular complexity analysis. The process is illustrated in Figure~\ref{fig:case_study_process}.

\begin{figure}
\centering
\includegraphics[width=0.92\textwidth]{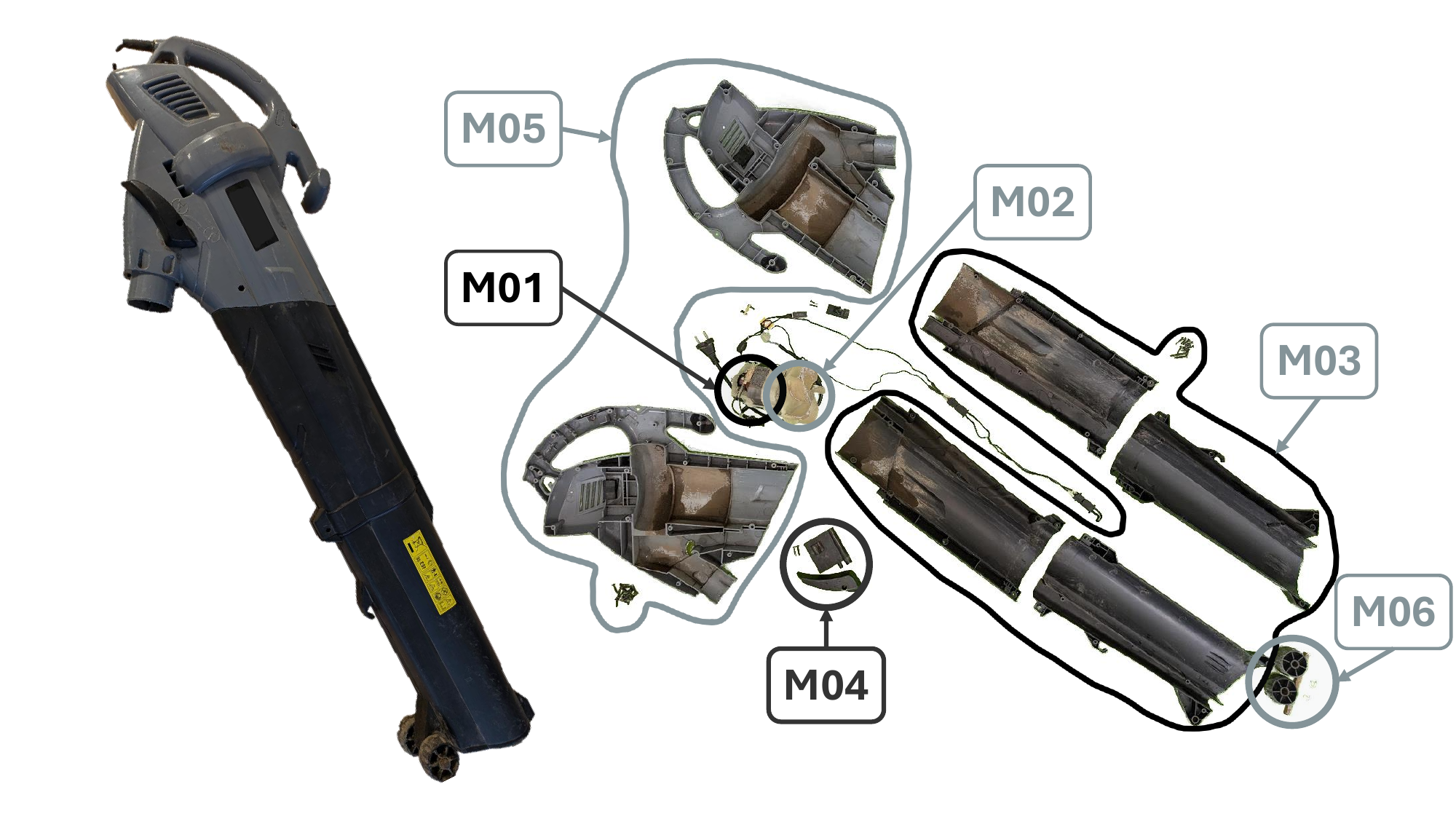}
\caption{\textbf{Exploded view of the handheld leaf blower baseline product.} Six distinct modules are represented:~M01 Motor, M02 Air generation (not disassembled for simplicity), M03 Air delivery, M04 Control, M05 Housing, and M06 Mobility.}
\label{fig:leaf_blower_photo}
\end{figure}

Through functional analysis, disassembly, and earlier workshop studies~\cite{monetti_evaluating_2025}, the product was decomposed into six primary modules.
\begin{itemize}
    \item[\textbf{M01}] \textbf{Motor module} — electric motor and plug-in interface;
    \item[\textbf{M02}] \textbf{Air generation module} — fan system generating airflow;
    \item[\textbf{M03}] \textbf{Air delivery module} — nozzle and outlet tube;
    \item[\textbf{M04}] \textbf{Control module} — on/off switch and speed selector;
    \item[\textbf{M05}] \textbf{Housing module} — structural plastic shell;
    \item[\textbf{M06}] \textbf{Mobility module} — wheel attachment for ground support.
\end{itemize}
A photograph of the product is shown in Figure~\ref{fig:leaf_blower_photo}, where modules are annotated based on their physical and functional roles. Structurally, the housing module (M05) serves as the assembly base. M01 (motor), M02 (fan), and M04 (control) are inserted into it. M03 (nozzle) and M06 (wheels) are attached in later stages. Full enclosure requires nine screws, and wiring between M04 and M01 must be routed manually through narrow internal passages.

Eight module-to-module interfaces were identified, both mechanical and electrical. These included a motion-transfer interface (M01–M02), several structural joints (e.g., M01–M05, M03–M05), and electrical linkages (e.g., M01–M04). Key assembly challenges included accurate shaft alignment between M01 and M02, difficult wiring between M04 and M01, and angular insertion of multiple modules, limiting gravity-aided assembly and increasing ergonomic strain.

The baseline configuration was used as a reference architecture for the following re-design steps, including:~evaluating technical solutions using the DFA-informed concept evaluation criteria in the DPM; re-assessing module boundaries using MIM with assembly-relevant drivers; visualising and classifying interfaces using the ADCD; quantifying assembly challenges through MSASM and TAC metrics.The baseline therefore serves both as a reference architecture and as an illustrative case for demonstrating the step-by-step application of the method and its capacity to generate actionable design improvements.

\subsection{Functional decomposition and concept evaluation}
To initiate the redesign process, a full functional decomposition of the leaf blower was conducted. This identified 18 functions across three categories:~core operational tasks (e.g., generating and directing airflow), auxiliary features (e.g., handling, control), and production-related needs (e.g., power supply, structural support). For simplicity, three alternative TS were considered per function.

To support early concept selection, four DFA-based criteria were used to evaluate each TS (see Table~\ref{tab:dfa_criteria}):~ease of assembly, scalability opportunities, automation potential, and support for late configuration. These reflect the strategic ambitions of the case company:~enabling low-cost assembly with high robustness, and supporting product variety without compromising manufacturability. Each TS was scored on a triadic ordinal scale (9 = favourable, 3 = moderate, 1 = unfavourable). While no numerical weights were applied, evaluators documented qualitative justifications to ensure traceability.

For the function \textit{Protect internal components}, three TS were compared: internal shock mounts (TS1), a reinforced motor casing (TS2), and a sealed electronics housing (TS3). TS2 was preferred due to its geometric simplicity and compatibility with automation. Although TS3 offered superior environmental protection, its reliance on adhesives and sealing steps made it unsuitable for early-stage automation or fast assembly. For \textit{Blow leaves and debris}, a centrifugal impeller was selected over axial or turbo-style fans. The impeller could be pre-tested with the motor and pre-assembled into a sub-module, supporting plug-in integration and scalable production. Similarly, the function \textit{Provide adjustable airflow} was implemented using a variable speed trigger instead of rotary knobs or digital control. The trigger offered direct user feedback and minimal interface complexity, reducing both ergonomic load and insertion requirements.

Across all functions, the same logic applied:~TS were favoured when they enabled pre-assembly, reduced reorientation or tool use, and supported intuitive interface alignment. This structured assessment made downstream module grouping more deliberate and aligned with assembly feasibility. Although the process was qualitative, it allowed for systematic exclusion of low-feasibility options early on, even without detailed CAD models. This helped refine the solution space while remaining lightweight and iteration-friendly. The outcomes of this stage directly informed the MIM and supported module formation.

\subsection{Module structuring using assembly-oriented drivers} \label{sec:mim_case}
Following concept evaluation, the selected TS were grouped into four final modules using both traditional drivers and newly introduced assembly-oriented drivers. This is done in the MIM, where architecture formation is guided by functionality and by assembly feasibility. The resulting architecture is a refinement of the legacy design, reducing module number and interface complexity.

The updated modular structure consists of the following four modules.
\newline
\textbf{MFU — Motor–Fan Unit}
A compact subassembly integrating the centrifugal impeller with a reinforced electric motor casing. This integration eliminates the need for precision shaft alignment and dedicated fixtures, enabling vertical insertion and simplified housing integration. The MFU is designed as a sealed, pre-tested unit.
\newline
\textbf{MAD — Module Air Delivery}
A redesigned airflow system that includes a telescopic nozzle and a built-in ground transport mechanism. This one-piece solution replaces the former split between nozzle and wheel attachment, reducing fastener use and removing non-essential components. The integrated design improves ergonomics and user handling while eliminating a module (Mobility) that did not contribute clear functional value for the target user segment.
\newline
\textbf{MPC — Module Power/Control}
A module that integrates variable-speed trigger control, start button, and a heavy-duty power cable with internal safety electronics. Pre-wired and pre-assembled, the MPC reduces manual routing, avoids tight internal passages, and enables plug-type connections to the MFU. This modularisation also supports straightforward replacement or servicing of control components, improving maintainability.
\newline
\textbf{MHO — Module Housing/Ergonomics}
A high-impact ABS clamshell casing that also incorporates structural and ergonomic functions. Snap-fit mechanisms replace torque-based fasteners, allowing single-handed insertion and improving self-alignment. Rubberised grip areas and a centralised handle structure enhance comfort and usability. This module serves as the structural base for assembly and acts as the interface hub for all other modules.

This architecture reflects the rethinking of the legacy module boundaries based on DFA-informed considerations. The total module count was reduced from six to four, simplifying both physical integration and production planning. High-complexity interfaces such as manual wiring, shaft alignment, and tool-based fixturing were eliminated. Pre-assembly opportunities were expanded through modular integration of control and power systems (MPC) and mechanical functions (MFU and MAD).

However, these improvements involved trade-offs. The unification of fan and motor into a sealed MFU limits upgrade flexibility, while the removal of wheel supports may reduce suitability for users with mobility or terrain-related needs. These decisions show some of the value of structured driver application:~they do not enforce one `correct' architecture, but enable reasoned evaluation of trade-offs between performance, functionality, and production strategy.

This outcome illustrates how assembly-oriented module drivers can improve architectural decision-making by proposing manufacturability early in the process. This helps reducing downstream effort while still preserving alignment with product functionality and user expectations.

\subsection{Interface analysis and complexity classification}
Interface analysis was performed using the taxonomy and coding scheme introduced in Section~\ref{sec:mfd}. Each connection between modules was classified according to interface type, task-level complexity, and assembly priority. Where multiple interface types coexist, these are coded separately and presented jointly.

In the legacy architecture, nine interfaces were identified. Notably, M01–M02 formed a motion-transmitting shaft interface, while M01–M04 combined electrical power and embedded control signals. Multiple interfaces required multi-axis reorientation and precise alignment, limiting automation and ergonomy.

The new four-module architecture reduced the number of interfaces to six and restructured several key connections. The shaft interface between motor and fan was eliminated and integrated in pre-assembly processes via module integration. Control and power systems were united within a pre-wired Module Power/Control (MPC), simplifying electrical routing and minimising manual threading. Previously embedded control wiring is now replaced with plug-type connectors and pass-throughs, improving accessibility and decoupling signal paths. The air delivery system (MAD) is now self-contained, eliminating the previous wheel attachment interface. Table~\ref{tab:interface_comparison} summarises the interface classification before and after modular restructuring.

These interface codes feed directly into the ADCD (Section~\ref{subsec:adcd}) and MSASM (Section~\ref{subsec:msasm}), linking task-level complexity to layout and strategy. This highlights the benefit of early interface modelling not just for functional clarity but also for guiding architectural simplification, ergonomic improvement, and automation readiness.

\begin{table}
\tbl{Interface classification before and after modular restructuring.}
{\begin{tabular}{>{\raggedright\arraybackslash}p{0.10\linewidth}
>{\raggedright\arraybackslash}p{0.15\linewidth}
>{\raggedright\arraybackslash}p{0.30\linewidth}
>{\raggedright\arraybackslash}p{0.35\linewidth}}
\toprule
\textbf{Module pair} & \textbf{Interface type} & \textbf{Interface code(s)} & \textbf{Comment} \\
\midrule
\multicolumn{4}{l}{\textit{Legacy architecture}} \\
\midrule
M01–M02 & M1 (shaft) & \texttt{M1-P3-H1-A2-T1-I0-D0} & Precision shaft alignment \\
M01–M04 & E+C & \texttt{E-P1-H1-A1-T0-I2-D1} \texttt{C-P1-H0-A0-T0-I0-D0} & Power and control wiring \\
M01–M05 & M2 & \texttt{M2-P2-H2-A1-T1-I2-D1} & Screw-mounted structural base \\
M02–M05 & M2 & \texttt{M2-P2-H2-A1-T1-I1-D1} & Fan-to-housing fastening \\
M03–M05 & M2 & \texttt{M2-P2-H2-A1-T0-I1-D0} & Snap or push-fit nozzle \\
M03–M06 & M2 & \texttt{M2-P1-H2-A3-T1-I1-D1} & Wheel mount with guided screws \\
M04–M05 & M2 & \texttt{M2-P2-H2-A1-T1-I1-D1} & Lateral mounting for control \\
M03–M04 & C & \texttt{C-P1-H0-A0-T0-I0-D0} & Safety interlock between control and power \\
\midrule
\multicolumn{4}{l}{\textit{DFA-informed architecture}} \\
\midrule
MFU–MHO & M2 & \texttt{M2-P2-H2-A1-T1-I1-D1} & MFU inserted top-down into housing \\
MFU–MPC & E+C & \texttt{E-P1-H1-A1-T0-I0-D0} \texttt{C-P1-H0-A0-T0-I0-D0} & Embedded signal and power connection \\
MAD–MPC & C & \texttt{C-P1-H0-A0-T0-I0-D0} & Safety interlock on air delivery \\
MAD–MHO & M2 & \texttt{M2-P2-H2-A1-T0-I1-D0} & Snap-fit nozzle into housing \\
MPC–MHO & M2 & \texttt{M2-P2-H2-A1-T0-I1-D0} & Lateral mounting for power/control unit \\
\bottomrule
\end{tabular}}
\label{tab:interface_comparison}
\end{table}

\subsection{Assembly Directions and Connections Draft} \label{subsec:adcd}
To translate the interface analysis and modular restructuring into spatial and procedural logic, a revised ADCD was developed. This diagram supports conceptual modularisation by combining interface modelling with early-stage assembly layout planning. It visualises insertion directions, interface types, to support ergonomic feasibility and automation readiness.

Unlike the legacy configuration where modules were inserted from multiple directions and required reorientations, the new ADCD has a more coherent assembly flow. The MHO acts as the structural base, into which all other modules are inserted. The MFU is inserted vertically from above, with gravity assistance and self-location. The elimination of the interface and improved fixture alignment supports single-operator assembly and better automation readiness. The MPC is mounted laterally into MHO via a standardised plug-type electrical interface and embeds control signal pass-through. Both connections reduce the need for manual wiring, removing internal routing paths. Additional M2-type mechanical interfaces secure the MPC to the housing, facilitating easy insertion and removal. The MAD, which includes a curved nozzle and integrated ground support, connects frontally to the housing via a tool-less M2-type interface. An additional control interface  links MAD and MPC to ensure safety interlock:~if the nozzle is removed, the power module is disabled automatically.

\begin{figure}
    \centering
    \subfigure[\textbf{Legacy architecture:} the ADCD diagram for the original leaf blower. Coloured shell layers distinguish inner modules from external add-ons. Labels highlight areas of ergonomic difficulty, reorientation, or manual wiring.]{
        \includegraphics[width=0.7\textwidth]{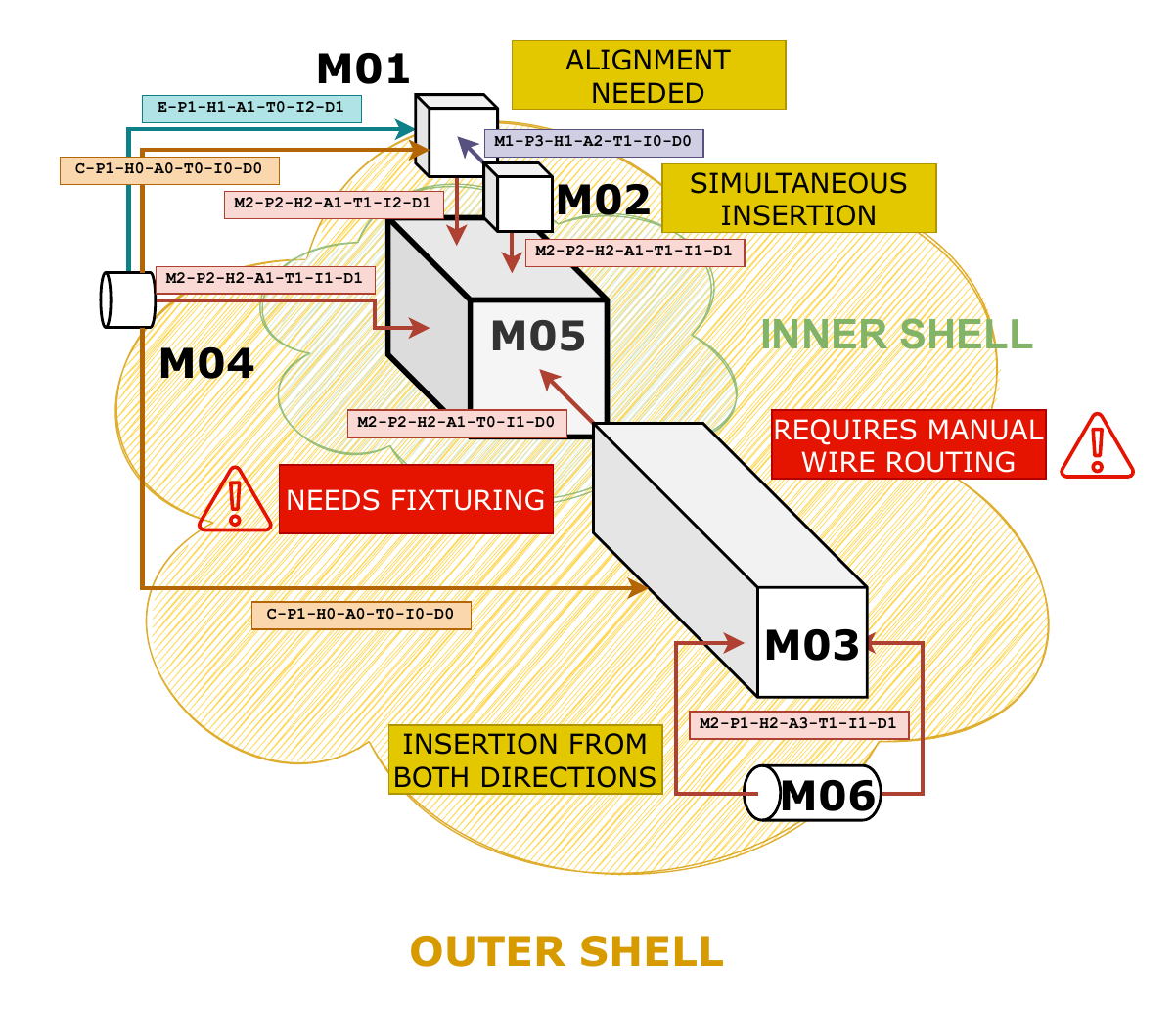}
        \label{fig:adcd_legacy}
    }
    \vfill
    \subfigure[\textbf{DFA-informed architecture:} the revised ADCD layout simplifies spatial logic with top-down and radial insertions into a central housing fixture. Reduced reorientation and clearer insertion vectors reflect improved assembly ergonomics and automation readiness.]{
        \includegraphics[width=0.7\textwidth]{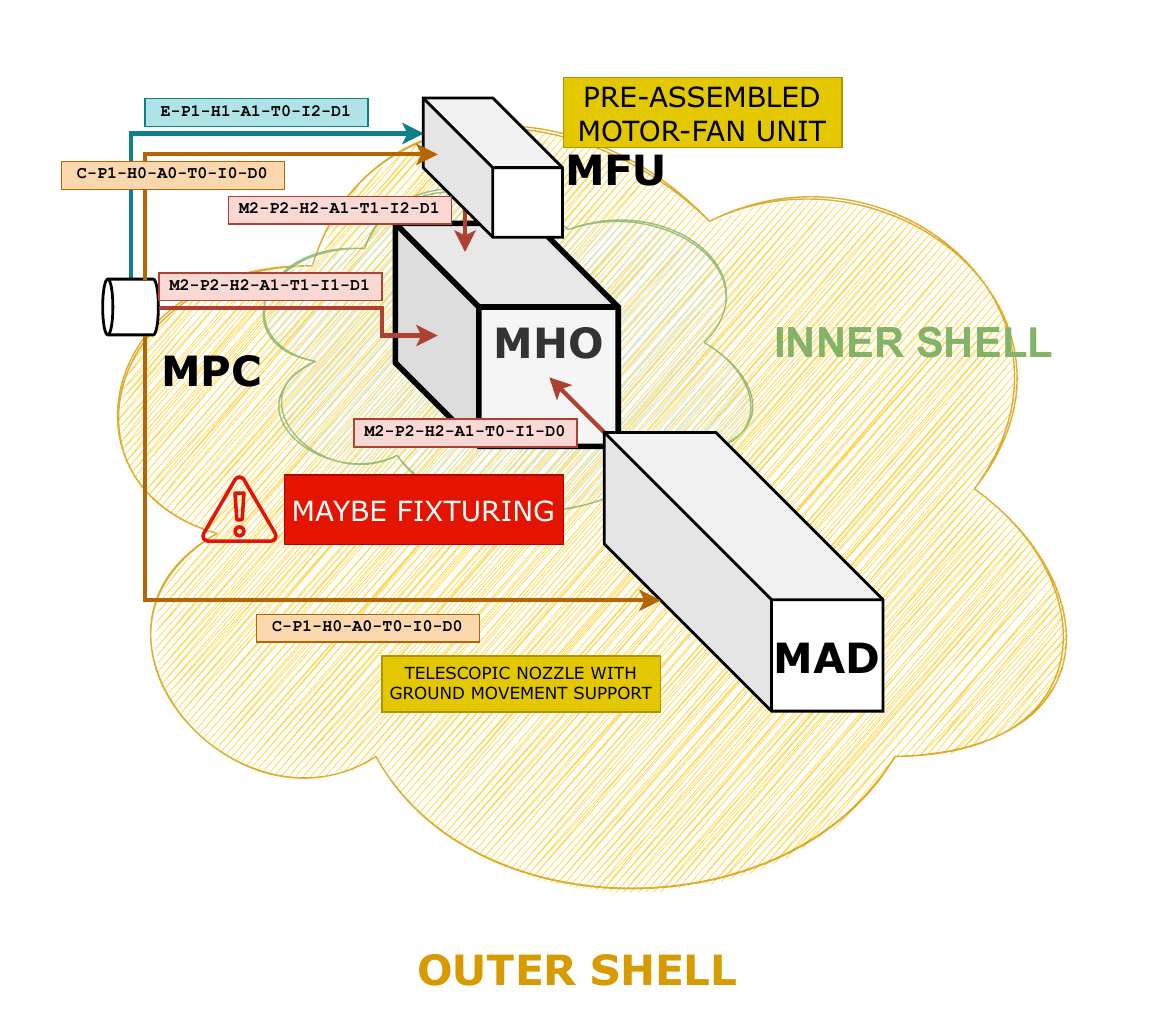}
        \label{fig:adcd_dfa_layout}
    }
    \caption{\textbf{ADCD diagrams for the leaf blower case.} Comparison before and after modular restructuring.}
    \label{fig:adcd_comparison}
\end{figure}

Figure~\ref{fig:adcd_comparison} presents the legacy architecture (Figure~\ref{fig:adcd_legacy}) and the revised architecture (Figure~\ref{fig:adcd_dfa_layout}) ADCD layout. Insertion arrows indicate directional vectors—top-down, lateral, or frontal, and are colour-coded by interface type. Shell layers distinguish internal modules from external ones.

Beyond visualisation, the ADCD enables early identification of critical assembly factors:~possible tool conflicts, reorientation steps requiring fixtures or jigs, parallelisation opportunities in sub-assembly, and ergonomic or automation constraints. It also supports mapping of interface complexity, task-level risks, and sequencing logic in downstream planning.

Overall, the revised architecture presents advantages over the legacy. Insertions now follow a consistent vertical or radial pattern, reducing spatial and ergonomic complexity. Module reorientation is no longer required, allowing simpler fixture design. The use of standardised interfaces reduces part variability, while eliminating fasteners shortens cycle time and simplifies tasks. Manual wiring and housing closure operations have been minimised. Gravity-assisted assembly and plug-in connections enhance both reliability and automation feasibility. The result is a compact, assembly-oriented architecture with fewer interfaces, clearer spatial logic, and lower overall complexity.

\subsection{Module Set Assembly Strategy Matrix and complexity metrics} \label{subsec:msasm}
The MSASM supports early-stage evaluation of connections by scoring each module set across 13 criteria, weighted according to production strategy (see Section~\ref{sec:msasm}). Without requiring geometric modelling, the matrix relies on qualitative inputs to assess the assembly feasibility and strategic desirability of each modular connection.

In the revised architecture, six module sets were evaluated using the same weighting profile as the legacy system. The results below reflect weighted averages across all criteria.
\newline
\textbf{MFU–MHO} (gravity-assisted insertion into housing): 7.65 — \textit{Good};
\newline
\textbf{MFU–MPC} (electrical and control connector): 5.84 — \textit{Caution};
\newline
\textbf{MAD–MHO} (tool-less snap-fit to housing): 8.11 — \textit{Good};
\newline
\textbf{MPC–MHO} (mechanical fastening with pass-through wiring): 7.11 — \textit{Acceptable};
\newline
\textbf{MAD–MPC} (safety interlock signal interface): 7.76 — \textit{Acceptable}.

While most connections were rated as \textit{Good} or \textit{Acceptable}, the \textbf{[MFU–MPC]} set scored below $6.0$ and was flagged for caution. This score remains within tolerable limits for refinement rather than full redesign, and reflects the presence of electrical contact tolerances and non-gravity-aided insertion.

The average MSASM score increased from $5.2$ in the legacy architecture to $7.3$ after modular restructuring, showing improvement in assembly feasibility. This was driven by the replacement of high-complexity interfaces (e.g., shaft couplings, embedded control wiring) with gravity-aided insertions and plug-style connectors. For example, the \textbf{[MFU–MHO]} set benefited from reduced alignment constraints and fewer tooling requirements, while \textbf{[MFU–MPC]} replaced cable threading with pre-wired snap connections.

To further complement the MSASM, Total Assembly Complexity (TAC) was computed. For the revised architecture, the five module sets included the following operations and scores.
\newline
$TAC_\text{MFU-MHO} = C_t = 1.50$;
\newline
$TAC_\text{MFU-MPC} = T02 + T03 = 0.75 + 0.75 = 1.50$;
\newline
$TAC_\text{MAD-MHO}	= 1.00$;
\newline
$TAC_\text{MPC-MHO}	= 2.50$;
\newline
$TAC_\text{MAD-MPC}	= 0.75$.
\begin{equation}
    TAC_{\text{architecture}} = \sum_{j=1}^{m} TAC_{\text{module set}_j} = 7.25.
\end{equation}
The total TAC for the revised design is $7.25$, down from $26.25$ in the legacy architecture—a 71~\% reduction. No single operation exceeds a $C_t$ of $2.5$, and four out of six operations score below $1.5$. This reduction is attributed to the elimination of complex shaft couplings, cable threading, and multi-handed assembly tasks present in the previous design.

These changes resulted in a more automation-friendly and ergonomically robust architecture, with reduced operator strain and clearer sequencing. The MSASM enabled early identification of high-effort interfaces, prioritised key redesigns, and guided module formation towards reliable, scalable connections. While further industrial validation is encouraged, the case confirms the practical value of integrating DFA-based complexity heuristics into early modular architecture development.
\section{Discussion} \label{sec:discussion}
This section discusses the findings and their contribution to the research questions. The results highlight potential to streamline integration, reduce operator workload, and improve automation compatibility. Although focused on architectural refinement rather than full redesign, the study demonstrates the full methodological flow, from function mapping to interface classification and module set evaluation.

Overall, the expanded MFD shifts DFA from a post-hoc synthesis tool to a strategic analytical method, embedding assembly thinking upstream. It allows cross-functional teams to shape product architecture with DFA and promotes early discussion of trade-offs before modules are fully defined.

\begin{itemize}
    \item[\textbf{RQ1}] was addressed by proposing an expanded MFD method that integrates DFA principles via new design tools and visual supports.
    \item[\textbf{RQ2}] was tackled through the introduction of predefined DFA-oriented evaluation criteria and assembly-oriented MDs, used in concept evaluation and module clustering.
    \item[\textbf{RQ3}] was supported by the implementation of quantitative assembly assessment using the weighted MSASM scores and the TAC metric.
    \item[\textbf{RQ4}] was answered through the adoption of an extended interface taxonomy that supports strategic evaluation of interface complexity and attachment logic.
\end{itemize}

\subsection{Contributions to addressing the research gap}
The expanded MFD method addresses a key gap in modular product development:~the absence of systematic assembly integration in early design.

First, DFA principles are embedded in two phases of conceptual development: (i) TS selection, via structured evaluation heuristics in concept evaluation; and (ii) module grouping, through new assembly-oriented MDs. This ensures alternatives are assessed not only for functional and market fit, but also for handling effort, reorientation difficulty, fastening strategy, and pre-assembly potential. Assembly thus shifts from a downstream constraint to an explicit design objective, moving DFA from reactive local optimisation to proactive architectural shaping.

Second, decision-support tools strengthen evaluation. The ADCD visualises sequencing and spatial logic, while the MSASM provides structured assessment of module set performance. Both replace ad hoc judgement with traceable, repeatable reasoning, improving confidence in early decisions.

Third, complementary metrics support strategic and operational assessment. The MSASM indicates compatibility with manual and automated assembly through weighted criteria, while the $TAC$ metric quantifies task complexity via interface coding. Linking physical, electrical, and other interfaces to task complexity makes them central to modular evaluation. Though not a substitute for geometry-based DFA, these metrics give actionable indicators of assembly effort in concept development, enabling early iteration and risk assessment.

Together, these contributions answer calls for a more integrated DFX–modularisation approach, aligning architectural reasoning with production strategy from the outset. The method enhances MFD’s evaluative capacity without altering its core structure, making it suitable for firms pursuing automation, simplified assembly, or scalable platforms.

It balances added modelling effort with tangible benefits. While heuristics, matrices, and complexity scores increase analytical workload, this can be offset by reduced late-stage redesign. Architectures optimised for handling, reorientation, and interface standardisation are more robust to manufacturing variation and operational error. Early alignment with these principles may also improve scalability, lower labour costs, and strengthen communication across design, production, and supply chain teams.

The proposal is also flexible. For instance, the MSASM weighting scheme lets firms emphasise criteria aligned with their production goals—labour reduction, automation readiness, or rapid reconfiguration. Rather than prescribing outcomes, the method enables structured, transparent decision-making from early design.

\subsection{Interpretation of case study results}
Despite its moderate complexity, the leaf blower exemplified design challenges typical of consumer-grade modular systems:~compact packaging, dense internal wiring, and limited accessibility during assembly. These constraints translated into concrete architectural trade-offs between structural simplicity, part accessibility, and interface routing—offering a useful test case.

The legacy architecture showed a loosely structured decomposition, with several modules connected through ad-hoc wiring and complex insertion sequences with reorientation. Applying the expanded MFD method led to the identification and rationalisation of these inefficiencies. Grouping the motor and fan into a pre-tested MFU reduced the number of exposed high-priority interfaces during final assembly, concentrating complex operations in upstream submodules. This enabled a shift toward a base-first stacking strategy, streamlining sequencing and improving fixture compatibility.

The MSASM scores flagged problematic module sets, such as motor–control and motor–fan, due to unfavourable insertion directions and handling burdens. The revised module sets scored significantly higher, showing promise in the use of early-stage heuristic evaluation for architectural improvement.

Similarly, the TAC metric revealed critical task-level burdens in the original layout, notably manual cable routing and fastener insertion within constrained cavities. These points were addressed by simplifying interfaces, adopting plug-and-play connectors, and relocating interaction points for improved accessibility. The resulting ADCD diagram visualised a more linear and coherent assembly flow, free of reorientation and compatible with gravity-assisted operations.

The case study shows the method’s practical usability. All tools were applied without requiring high-fidelity modelling. This reinforces the idea that early DFA integration need not depend on resource-heavy tools but can instead be embedded into standard concept development workflows.

One potential crucial outcome was that the team working on the project was allowed to explicitly visualise and weigh trade-offs—between handling complexity and sub-assembly integration; between interface simplification and module interchangeability; and between automation compatibility and functional proximity. These trade-offs are often negotiated tacitly in practice, but the method showed them systematically and early. In this way, the case illustrates how lightweight but structured evaluation tools can support robust decision-making mediating between functionality, manufacturability, and flexibility.

Notably, the decomposition supported the identification of a `shell logic' where inner modules concentrate core product functions and outer modules accommodate customisable features or product differentiation. This spatial-functional separation helps streamlining sequences and interface management but also supports CE objectives. Outer modules associated with styling or end-user features can be more readily disassembled, reused, or upgraded, reducing lifecycle waste and enabling service-friendly design strategies. Such alignment between architecture and sustainability objectives is often difficult to achieve without early-stage structural reasoning.

The case also suggests that the method is particularly effective in brownfield scenarios, where an architecture is established and assembly-related might be easily identified. In such contexts, it provides a practical means to rationalise module boundaries, simplify interfaces, and reduce handling complexity without radical overhauls. Nonetheless, it could be equally suited to greenfield development, where it can structure early architectural decisions before any physical embodiment is fixed. Its capacity to guide design based on production constraints, even in the absence of legacy artefacts, shows flexibility across scenarios.

\subsection{Reflections on method feasibility and scalability}
From a feasibility standpoint, the proposed method has been deliberately designed to preserve compatibility with the established MFD workflow, without disrupting its core logic or industrial uptake. The enhancements can be implemented using standard tools such as spreadsheets and worksheets, requiring no advanced software or modelling. The case study highlights that, even with moderate effort, these additions provided actionable insights and supported early-stage architectural refinement.

Structurally, it is scalable as its criteria and flexible weighting system allow to adapt it to design contexts and strategic priorities. Companies aiming for automation, for instance, can assign higher weights to MSASM criteria related to feeding direction, or grasping principles. Others may emphasise ergonomic handling, or end-of-life disassembly. This adjustability supports incremental optimisation and more ambitious modular redesigns, to accommodate varying industrial objectives and maturity levels.

Nonetheless, inherent limitations remain. Subjective judgement plays a central role in several steps, particularly applying heuristic criteria, estimating task complexity, and assigning interface priorities. While this is consistent with early-stage design practices, it introduces variability, particularly among less experienced teams or cross-functional groups unfamiliar with assembly analysis. On the other hand, this could serve as the platform for launching proactive and productive cross-functional collaborations. Care should be taken during the process, but establishing internal standard procedures, and using facilitation workshops, may help mitigate inconsistency in decision-making.

Scalability also has practical barriers. Broader deployment across entire product families or in platform development requires more extensive data collection, disciplined interface mapping, and coordinated team input. The workflow is not inherently data-heavy, but its effectiveness depends on sufficient granularity and shared understanding of interface logic. Digital integration with CAD environments or PLM systems could support these needs in more complex applications.

It is also worth noting that while the method supports radical architectural transformation, the case study focused on a moderate refinement. This was intentional, as it allowed demonstration of the full methodological workflow without overwhelming the reader with product-specific detail. However, future applications to more complex systems such as tightly integrated mechatronics or adaptive platform architectures may demand additional tooling for managing architectural dependencies and variant propagation.

Overall, a pragmatic yet powerful extension of MFD was built for early-stage modular product design. Its balance of structure and usability enables structured reasoning without excessive analytical burden, making it suitable for both industrial application and educational use. By embedding assembly-aware thinking into the early design space, it supports transparency, cross-functional communication, and downstream production alignment at a point where change is still affordable.

\subsection{Implications for industry practice}
The proposed method addresses one of the issues of industrial practice:~costly late-stage rework caused by insufficient consideration of manufacturability and assembly during concept selection and module definition. By embedding DFA principles, manufacturers can anticipate production challenges, select friendly technical solutions, and reduce the risk of downstream inefficiencies. Moreover, the inclusion automation-relevant criteria supports strategic alignment for firms aiming to transition toward higher levels of automated or semi-automated assembly. Rather than treating automation as a post-hoc optimisation, reducing reliance on manual labour and complex fixtures is encouraged at the outset.

Standardised interfaces and rational module boundaries also facilitate future variant generation and platform scalability, while reducing risks from change propagation. These benefits can translate into lower lifecycle costs, better engineering change processes, and more resilient supply chains.

From a practical standpoint, a modest increase in analytical effort might be introduced, through a preliminary function analysis, definition of technical alternatives, and identification of key interfaces, activities that are typically present in existing modularisation routines. The proposed tools increase focus toward assembly feasibility and strategic alignment, rather than imposing new modelling burdens.

It should be noted that the case study was conducted outside an industrial setting. While the redesign logic and evaluation tools were based on real-world practices, no empirical time or cost measurements were taken. As such, while high in conceptual coherence and practical usability, further validation in pilot projects or industrial trials is necessary to confirm quantitative performance impacts such as assembly time, quality yield, or cost-per-unit.

Nonetheless, the outcomes suggest that the method holds strong potential as a decision-support tool for companies seeking to bridge design and operations through modularisation. By bringing production constraints and opportunities into the early design space, it enables proactive rather than reactive development.

\subsection{Future work directions}
While the proposed method shows conceptual coherence and internal consistency, further research is needed to expand its scope, improve its rigour, and explore its integration into industrial contexts.

A key next step is validation through real-world implementation. Test-cases in ongoing modularisation projects to evaluate practical utility and scalability should be run. Such applications would allow assessment of the  impact on assembly lead time, iteration cycles, and cross-functional collaboration. Comparative studies across product generations or platforms could help evaluate its effectiveness in reducing rework and enhancing alignment between design and production strategy.

Tool integration and workflow compatibility also merit further development. Although the method has been designed to align with standard MFD platforms and existing modularisation routines, certain steps could benefit from partial automation. CAD-based interface detection, automated extraction of mating orientations, and integration with assembly simulation tools could accelerate analysis and reduce designer workload. Similarly, better connection with configuration platforms and PLM systems could enhance usability in large-scale variant planning contexts.

Opportunities for simplification and extension also exist. While the full MSASM provides a comprehensive perspective, streamlined versions could be developed for rapid-decision making. These might reduce the number of criteria, apply binary scoring, or focus only on automation-critical aspects. Likewise, the output could be extended with automated flagging of complexity risk patterns or penalty rules for recurring constraints, supporting escalation procedures during reviews. In the criteria-based concept selection of TS lies a low-hanging fruit possibility for the expansion of heuristics based on other relevant DFX methodologies, focused for example on disassembly principles, recycling opportunities, or life-cycle assessment. This could easily enhance the capabilities of the tool to reason within the CE framework, and help promote more sustainable products.

Finally, future research could support semi-quantitative calibration of weights and scores. While it currently relies on expert-defined weightings and ordinal scales, industrial datasets and best-practice repositories could inform more data-driven evaluations. These could be used to refine score thresholds, establish industry-specific scoring templates, or enable predictive assessments of assemblability.
\section{Conclusion} \label{sec:conclusion}
Recognising that traditional modularisation workflows tend to consider assembly primarily as a downstream optimisation activity, this study set out to develop structured, analytically grounded tools that support modularisation from the outset of product architecture development. The goal was targeted by positioning DFA not as a corrective measure, but back in its rightful position as a foundational rationale that can shape design from the outset.

The main contribution is the development of an expanded MFD method that embeds this logic directly into modular concept generation and evaluation. By introducing assembly-oriented heuristics, a structured interface taxonomy, and evaluation tools, it enables early alignment between architecture and production strategy. It maintains compatibility with the established process while significantly extending its decision-making power.

The illustrative case study showed potential to reduce complexity, streamline sequencing, and highlight trade-offs. Tools such as the ADCD supported visual reasoning and early identification of criticalities, while MSASM facilitated comparing module sets. These additions proved useful even in a brownfield scenario, supporting refinement of an existing product without requiring full rearchitecting.

What distinguishes this work is the shift in perspective by treating assembly not as a constraint to be accommodated later, but as a strategic input to shape cleaner architectures, enable automation, and enhance lifecycle adaptability. Emergent insights highlight capacity to support circularity goals such as reuse, upgrade, and easier disassembly.

Researchers noted years ago that assembly considerations deserve a more central place in architectural thinking. When integrated proactively, they can guide cleaner modular configurations, reduce total assembly effort, and improve product adaptability. This alignment of design intent and operational logic is essential not only for cost control but also for enabling strategic objectives.

The next steps will focus on real-world validation in industrial settings, digital tool integration, and potential simplification for rapid application in variant planning or platform design. Equally important is the need for training and support material that can help engineering teams operationalise it with minimal uncertainties. Tailored implementation across products of varying complexity will also help refine score thresholds, scoring logic, and configurability patterns.

Ultimately, this work offers a bridge between conceptual modular architecture creation and production strategy, seen as a necessary step toward more robust and scalable modular systems. It invites both academics and practitioners to rethink the role of assembly in design as an enabler of architectural excellence. It re-establishes DFA as an analytical tool for early product development, complementing its traditional use for design refinement.

\section*{Acknowledgements}
The authors would like to thank Arne Erlandsson for the invaluable insights through the work. Their industrial experience has enriched the study and been source of inspiration.

\section*{Disclosure statement}
There are no competing interests to declare.

\section*{Funding details}
This work was supported by the European Union’s Digital Europe Programme grant number 101083691, part of the ShiftLabs initiative;and partially supported by XPRES Centre of Excellence in Production Research.

\section*{Data availability statement}
The data are available upon reasonable request.

\bibliographystyle{tfq} 
\bibliography{references}
\end{document}